\documentclass[fleqn,usenatbib]{mnras}
\usepackage{mathptmx}
\usepackage[T1]{fontenc}
\DeclareRobustCommand{\VAN}[3]{#2}
\let\VANthebibliography\thebibliography
\def\thebibliography{\DeclareRobustCommand{\VAN}[3]{##3}\VANthebibliography}
\usepackage{graphicx}	
\usepackage{amsmath}	
\usepackage{amssymb}	
\usepackage{rotating}
\usepackage{sidecap}
\usepackage{lscape}
\usepackage{multicol, blindtext}

\usepackage{aecompl}
\usepackage{wasysym}
\usepackage{array}
\usepackage{times}
\usepackage{url}
\usepackage{xcolor}
\usepackage{hyperref}
\hypersetup{colorlinks=true,citecolor=blue,linkcolor=blue,filecolor=blue,urlcolor=blue}
\usepackage{soul}

\title[Endothermic SIDM in Milky Way haloes]
    {Endothermic self-interacting dark matter in Milky Way-{like} dark matter haloes}
\author[S. O'Neil et al.]
    {Stephanie O'Neil$^{1}$\thanks{E-mail: sloneil@mit.edu},
    Mark Vogelsberger$^{1,2}$,
    Saniya Heeba$^{3}$,
    Katelin Schutz$^{3}$, 
    \newauthor
    Jonah C. Rose$^{4}$,
    Paul Torrey$^{4}$,
    Josh Borrow$^{1}$,
    Ryan Low$^{5}$,
    Rakshak Adhikari$^{5}$,
    \newauthor
    Mikhail~V. Medvedev$^{5,6}$,
    Tracy R. Slatyer$^{1,2,7}$,
    and Jes\'us Zavala$^{8}$
     \vspace{0.3cm}\\
     $^{1}$Department of Physics and Kavli Institute for Astrophysics and Space Research,
           Massachusetts Institute of Technology,
           Cambridge, MA 02139, USA\\
     $^{2}$The NSF AI Institute for Artificial Intelligence and Fundamental Interactions, Massachusetts Institute of Technology, Cambridge, MA 02139, USA\\
     $^{3}$Department of Physics \& McGill Space Institute, McGill University, Montr\'{e}al, QC H3A 2T8, Canada\\
     $^{4}$Department of Astronomy, University of Florida, Gainesville, FL 32611, USA\\
     $^{5}$Department of Physics and Astronomy, University of Kansas, Lawrence, KS 66045, USA\\
     $^{6}$Laboratory for Nuclear Science, Massachusetts Institute of Technology, Cambridge, MA 02139, USA\\
    $^{7}$Center for Theoretical Physics, Massachusetts Institute of Technology, Cambridge, MA 02139, USA\\
    $^{8}$Centre for Astrophysics and Cosmology, Science Institute, University of Iceland, Dunhagi 5, 107 Reykjavik, Iceland
    }

\begin{document}

\date{Accepted 2023 June 07. Received 2023 April 25; in original form 2022 October 27}

\pagerange{\pageref{firstpage}--\pageref{lastpage}}
\pubyear{2022}

\maketitle
\label{firstpage}

\begin{abstract}
Self-interacting dark matter (SIDM) offers the potential to mitigate some of the discrepancies between simulated cold dark matter (CDM) and observed galactic properties.
We introduce a physically motivated SIDM model to understand the effects of self interactions on the properties of Milky Way and dwarf galaxy sized haloes.
This model consists of dark matter with a nearly degenerate excited state, which allows for both elastic and inelastic scattering.
In particular, the model includes a significant probability for particles to up-scatter from the ground state to the excited state.
We simulate a suite of zoom-in Milky Way-sized \textit{N}-body haloes with six models with different scattering cross sections to study the effects of up-scattering in SIDM models.
We find that the up-scattering reaction greatly increases the central densities of the main halo through the loss of kinetic energy.
However, the physical model still results in significant coring due to the presence of elastic scattering and down-scattering.
These effects are not as apparent in the subhalo population compared to the main halo, but the number of subhaloes is reduced compared to CDM.
\end{abstract}


\section{Introduction}
\label{sec:intro}

Roughly $30\%$ of the Universe's energy budget is made of matter, of which roughly $15\%$ is baryonic \citep[e.g.][]{PlanckCollaborationXIII2016}.
The rest of the matter, dark matter (DM), is often modelled as a cold and collisionless substance (cold dark matter, CDM) that interacts solely through gravity. 
DM forms the backbone for structure formation, with galaxies forming in overdense regions seeded by dark matter.
\textit{N}-body simulations aiming to model structure formation on large scales have had significant success in reproducing key results like the distribution of galaxies and the halo mass function using abundance matching \citep[e.g.][]{Springel2008, BoylanKolchin2009, Klypin2011, Garrison-Kimmel2014a}.

Despite the many successes of these simulations, there are still several discrepancies between the standard CDM models and observations \citep[see][for review]{Bullock2017,Zavala2019a,Sales2022}.
Often noted is the ``missing satellites'' problem, where many more satellite haloes were found in simulations than observed satellite galaxies \citep{Moore1999,Springel2008,Klypin2011,Garrison-Kimmel2014a}.
However, this has been resolved through the detection of faint dwarf galaxies and accounting for the completeness of surveys bringing the number count for luminous galaxies into agreement with simulations \citep{Kim2018,DES2019}.
Abundance matching using the star formation rate rather than the stellar mass also adjusts the predicted halo mass function for larger subhaloes.
The expected galaxy count and larger dark matter halo masses, however, introduces an additional problem, the ``too-big-too-fail'' problem \citep{Boylan-Kolchin2011,Boyer2012,Garrison-Kimmel2014b}.
For dark haloes to exist and account for the missing satellites, especially for larger mass haloes, there must be haloes that have failed to form galaxies or been stripped of their stars despite their strong gravitational potential wells.

The central densities of dark matter haloes also tend to have steeper slopes in CDM \textit{N}-body simulations than those inferred from observations, known as the ``core-cusp problem'' \citep{Marchesini2002,deBlok2008,KuziodeNaray2008,Walker2011}.
While dark matter haloes in \textit{N}-body simulations can be universally modelled with the NFW profile, with an inner logarithmic slope near $-1$, many observed satellite galaxies tend to have flatter (cored) profiles with inner slopes between $0$ and $-0.5$.
Additionally, slopes of observed galaxy density profiles are not uniform as predicted by \textit{N}-body simulations.
The ``diversity of shapes problem'' refers to the observation that there is a diverse population of shapes of rotation curves for dwarf field galaxies \citep{Oman2015,Creasey2017} and inner density profiles of Milky Way satellites \citep{Zavala2019b,Hayashi2020}.
Observed satellite galaxies are also not distributed in the way predicted by CDM models and tend to be distributed along a plane rather than isotropically \citep{Kunkel1976,Lynden-Bell1976,Parlowski2015a}, though the extent to which this poses a problem for CDM is uncertain \citep{Pham2022}.

Most of these discrepancies initially became apparent by comparing DM-only simulations to observations, and the addition of baryonic physics may mitigate some of these issues.
Supernova feedback or stellar winds, for example, can push dark matter outwards and create a core \citep{Navarro1996b,Governato2012,DiCintio2014a,Dicintio2014b}.
Although early work questioned the strength of feedback effects at the mass scales of dwarf galaxies \citep{Gnedin2002}, more recent work has successfully produced cores with various feedback prescriptions in a range of galaxy sizes \citep[e.g.][]{Pontzen2012,Keller2016,Burger2022}.
\citet{Hayashi2020} also showed that feedback based on nonspherical Jeans analysis produces significant coring for galaxies with a mass of roughly one hundredth the mass of its main halo.
At lower and higher masses, dwarf haloes in their model still produced cusps, showing a diversity of shapes within their halo population.
However, these effects are not sufficient as of yet to conclusively resolve the diversity in halo density profiles \citep{Santos-Santos2020}, nor are baryons able to affect the spatial distribution of satellites to cause them to lie on a plane \citep{Pawlowski2015b}.

Additionally, simulations use subgrid baryonic models, so they are not fully resolved and can vary significantly depending on the model parameters \citep{Vogelsberger2020}.
\citet{Munshi2019}, for example, showed that altering the subgrid star formation criteria can change the resulting dwarf and ultrafaint dwarf satellite populations by a factor of 2 to 5, and \citet{Agertz2016} demonstrated that star formation models can have a significant effect on the morphology of simulated galaxies.
\citet{Benitez-Llambay2019} showed that altering the gas density threshold for star formation affects whether a core is formed or not in dwarf galaxies, which is a key test in comparing simulations and observations.
The efficiency of star formation in a simulation can be the difference between matching observed galaxy morphology and producing unrealistic galaxies \citep{Agertz2015}.
Even changing the hyrdrodynamics solver used to model gas can change the properties of galaxies.
\citet{Torrey2012}, for example, showed that \textsc{Arepo}, a moving mesh scheme, produced larger disks than \textsc{Gadget}, a smoothed particle hydrodynamics scheme.
While baryonic physics still needs to be better understood and modelled and could potentially lead to new insights in these small-scale dark matter problems, it is worthwhile to explore alternative dark matter candidates.

Many DM candidates can affect the formation and properties of DM haloes (see \citealt{Bechtol2022} for a recent review).
For instance, due to free-streaming behaviour in the early Universe, warm dark matter (WDM; \citealt{Bode2000}) and related DM candidates like sterile neutrinos \citep{Dodelson1993,Shi1998} and DM produced by freeze-in \citep{Dvorkin2020} can suppress the formation of low-mass haloes and thus mitigate any potential problems with missing satellites \citep{Colin2000, Zavala2009,Lovell2012}.
However, there are strong constraints on WDM (and related models) from the Lyman-$\alpha$ forest \citep{Viel2013, Schneider2014a,Irsic2017a,Hooper2022}.
Moreover, other pressing issues like the core-cusp or too-big-to-fail problems are not entirely solved by WDM-like DM candidates, though WDM in combination with baryonic effects does alleviate them compared to CDM \citep[e.g.][]{Colin2008,Polisensky2015,Lovell2017}.
Free-streaming of these particles in the early universe is not sufficient to explain the late-time properties of the interiors of DM haloes.

Self-interacting dark matter (SIDM; \citealt{Spergel2000}) is a promising alternative DM candidate that can affect the distribution of DM inside haloes by transferring orbital energy between different parts of the halo. This tends to reduce the central density of DM haloes and flattens the density profile, thereby alleviating the core-cusp and too-big-to-fail problems; for a review see \citet{Tulin2017}.
\citet{Kaplinghat2019} noted that high concentration subhaloes have higher dark matter densities compared to low concentration subhaloes since they are more likely to undergo core collapse, which contributes to solving the too-big-to-fail and diversity of shapes problems.
This process does not produce significant difference from CDM models at dwarf galaxy scales for viable constant cross sections \citep{Zeng2022,Silverman2022}, so these models are constrained to velocity-dependent cross sections, where the scattering probability depends on the relative velocity of the particles \citep{Meshveliani2022}.
Recent work, for example \citet{Sameie2020}, has also shown that SIDM combined with tidal forces produces a wide range of density profiles, thus alleviating the diversity problem as well.
Although SIDM induces significant departures from CDM on small scales, the behaviour of SIDM and CDM are indistinguishable on large scales, consistent with observations \citet{Rocha2013, Vogelsberger2016,Sameie2019}. Although SIDM was initially introduced as a phenomenological device to address small-scale structure issues, it is naturally realised in broad classes of DM models including atomic DM \citep{Cline2013b,Cyr-Racine2012,Boddy2016}, composite DM \citep{Frandsen2011,Boddy2014,Chu2018b}, strongly interacting massive particles \citep{Hochberg2014a,Hochberg2014b,Bernal2015,Choi2017,Hochberg2018b}, charged DM \citep{McDermott2010,Liu2019,Dvorkin2019}, mirror DM~\citep{Mohapatra2001,Foot2008,An2010,Chacko2021}, and secluded DM~\citep{Boehm2003b,Pospelov2007,Feng2008}, among many others.

Earlier SIDM simulation work focused primarily on elastic scattering with a constant (velocity-independent) cross section per mass ($\sigma_{\rm T}/m$).
These studies generally find that small $\sigma_{\rm T}/m\sim1\ \rm{cm}^2\rm{g}^{-1}$ are needed to match observations of galaxy clusters \citep{Rocha2013,Andrade2022} but larger cross sections are necessary to alter the inner densities and abundances of dwarf galaxies \citep{Zavala2013,Elbert2015}.
However, many of the model realisations of SIDM naturally predict velocity-dependent interaction rates, for instance due to the presence of a long-range mediator.
Such velocity dependence also arises for e.g. Rutherford scattering in the Standard Model (SM).
Velocity-dependent $\sigma/m$ could allow for significant scattering effects on small scales without violating constraints on cluster scales \citep{Loeb2011,vandenAarssen2012,Tulin2013,Todoroki+2019a,Todoroki+2019b,Todoroki+2022,Silverman2022}.
In the case of DM-SM scattering, it is possible to differentiate constant and velocity dependent $\sigma/m$ through detection signals in direct detection experiments \citep[e.g.][]{Rahimi2022}.
The recoil speed of particles in the detector, for example, could increase or the phase of the electric signal could shift, with constant $\sigma/m$ producing larger changes compared to CDM \citep{Vogelsberger2013b}.
The same force may mediate these interactions as DM-DM interactions, so potential direct detections of dark matter particles may be able to constrain $\sigma/m$.

In broad classes of DM models, DM interactions are inelastic \citep{mvm2000, mvm2001a, mvm2001b} and can cause qualitatively different behaviour compared to elastic SIDM \citep{Arkani-Hamed2008,mvm2010,mvm2010b,mvm2014PRL,mvm2014JCAP,Schutz2014,Blennow2016,Zhang2016,Vogelsberger2019,Alvarez2019}.
\citet{Alvarez2019} used a model with light dark force mediators to demonstrate that inelastic scattering leads to altered halo appearance and that the up-scattering reaction should not be significant in dwarf haloes to allow for halo formation.
Recently, the XENON1T experiment \citep{Aprile2005} 
observed an excess of electron recoil events at a few keV \citep{Aprile2020}, although this excess has not been seen by XENONnT \citep{Aprile2022}, pointing to likely tritium contamination.
The excess did, however, reinvigorate community interest in inelastic DM models that could explain the signal (see e.g. \citealt{Baryakhtar2020, Bloch2020, An2020}). It is therefore timely to consider how these models affect DM haloes (which, in turn, affects their predicted signals in direct detection experiments).

Inelastic DM models require an avenue for energy transfer to occur during the inelastic reactions.
One way to accomplish this is for DM particles to exist in multiple states with a mass difference, analogous to SM particles.
Exothermic reactions (e.g. scattering from the excited state to the ground state) can impart a large velocity kick to particles in the final state, which aids in the development of a core and makes it more difficult for subhaloes with small escape velocities to form. \citet{mvm2014PRL} showed that SIDM with inelastic (exothermic and endothermic) interactions with $\sigma_{\rm T}/m\sim 1\ \rm{cm}^2 \rm{g}^{-1}$ and the relative mass degeneracy between the DM states $\Delta m/m\sim \textrm{few}\times10^{-8}$ (corresponding to the characteristic kick velocity of about 100~km/s) can simultaneously create cored density distributions and affect the dwarf halo abundance.
These changes happen on small scales only, large scales remain unaffected and agree with CDM predictions.  
\citet{Vogelsberger2019} showed that inelastic reactions with cross sections per mass of the order of $\sigma_{\rm T}/m\sim0.1-1\ \rm{cm}^2/\rm{g}$ can create significant cores in dwarf galaxies and decrease their abundance.
\citet{Chua2021} noted that these reactions can core a main halo more efficiently than elastic collisions.
Other exothermic reactions like DM decay could also provide an avenue for large velocity kicks \citep{Wang2014}. \citet{Todoroki+2019a,Todoroki+2019b,Todoroki+2022} explored a large set of velocity-dependent and velocity-independent exothermic and endothermic reaction models across a halo mass range from dwarfs to clusters, $10^8M_\odot\lesssim M\lesssim 10^{15}M_\odot$. Their results agree with \citet{Vogelsberger2019} and favour the same range of cross-sections, as well as also indicate that the velocity-dependent models allow for a large diversity of DM density profiles. 
These previous simulations were initialised with high cosmological abundances of the excited DM state.
However, due to the high density of the early Universe and the fact that it is energetically favourable to down-scatter, the generic expectation is that the excited state will be cosmologically depleted. If the vast majority of the DM is in the ground state by the onset of structure formation, then exothermic reactions are not an available scattering channel. It may however be possible to have late-time \emph{endothermic} reactions in the right kinematic environment where DM in the ground state up-scatters into the excited state. It should be noted that there is an exception --- the SIDM with flavor mixing model called 2cDM --- that does not face the above described severe depletion of heavy excited states. In this scenario, such a depletion is avoided due to the coherent nature of inelastic scattering of particles in a flat universe before the structure formation, i.e., in the absence of large gravitational fields \citep{mvm2001a,mvm2010,mvm2014JCAP}.

Here, we perform the first simulations in which endothermic reactions dominate the particle interactions and the initial DM composition is comprised solely of the ground state (i.e. lowest mass) particles. These simulations are particularly relevant for the DM candidates that result in low populations of excited states in the early universe. In our simulations, endothermic reactions are kinematically accessible and can populate the excited state at late times. In contrast to exothermic reactions, endothermic reactions have the potential to \emph{increase} central halo densities due to the conversion of kinetic energy into mass. Our study complements previous simulations of DM with interactions that dissipate kinetic  energy. For instance, \citet{Shen2021,Shen2022} simulated a dissipative DM model in which particles lost a constant fraction of their kinetic energy when scattering and showed that dark matter profiles steepened and haloes contracted radially, though their mass remained similar to the CDM case.
\citet{Huo2020} compared a CDM \textit{N}-body simulation to a purely elastic model and several variations of dissipative cross sections and velocity thresholds, and they showed that dissipative dark matter can speed up gravothermal collapse and steepen the inner density profile.
\citet{Essig2019} similarly showed that dissipative dark matter can accelerate gravothermal collapse even with very long interaction timescales.

In this work, we introduce physically motivated endothermic SIDM models with parameters that yield a relic DM abundance from freeze-in that match the observed DM density. We simultaneously ensure that constraints on DM annihilation from the cosmic microwave background are satisfied as well as constraints on low-mass DM and the mediator from supernova~1987A and Big Bang Nucleosynthesis. Using a suite of dark matter-only zoom-in simulations, we explore the impact of this DM model on halo properties.
Using CDM as our fiducial model and the previously studied inelastic SIDM scenario (dominated by exothermic reactions) as a benchmark \citep{Vogelsberger2019}, we compare the main halo density profiles, subhalo abundance and densities. For SIDM models, we additionally track and compare the evolution of particle state fractions and reaction rates.

The rest of the paper is organised as follows.
In Section \ref{sec:methods}, we describe our simulations, SIDM model, and the implementation of this model.
We describe our results and analysis in Section \ref{sec:results}.
Finally, we summarise our conclusions in Section \ref{sec:conclusions}.

\section{Methods}
\label{sec:methods}

\subsection{Simulations}
\label{sec:methods_sims}
We run a suite of cosmological DM-only zoom-in simulations from the same initial conditions with a main halo of mass $M_{\rm200,mean}\approx2\times10^{12}$ M$_{\odot}$ at $z=0$, where $M_{\rm200,mean}$ refers to the mass within the radius where the average density is 200 times the mean density of the universe.
The simulations use cosmological parameters matter density fraction $\Omega_{\rm m} = 0.302$, dark energy fraction $\Omega_{\Lambda} = 0.698,$ and Hubble constant $H_0=100h\,\rm{km} \,\rm{s}^{-1}\,\rm{Mpc}^{-1}$ where $h=0.691$.
The simulation box has a side length of $100$ cMpc$/h$.
Each simulation is run from the same set of initial conditions that were generated with \textsc{Music} \citep{Hahn2011} at $z=127$.

There are 55,717,200 zoom particles that make up the main halo and its surroundings.
They have mass of $2.20\times10^5\ \rm{M}_{\odot}$ and a comoving softening length of $0.305$ kpc with a maximum physical softening length of $0.153$ kpc.
These particles are given additional stored properties associated with the self-interacting model (e.g. current state or number of scatters of the particle).
These particles are surrounded by four species of lower resolution background particles filling the rest of the simulation box with masses:
$2,081,478$ particles of mass $1.76\times10^6\ \rm{M}_{\odot}$,
$559,385$ of mass $1.41\times10^7\ \rm{M}_{\odot}$,
$223,154$ of mass $1.13\times10^8\ \rm{M}_{\odot}$ and
$134,167,340$ of mass $9.03\times10^8\ \rm{M}_{\odot}$.
The background particles have comoving and maximum physical softening lengths adjusted by a factor of the cube root of the ratio of the background and zoom particle masses.
We also run two lower resolution sets of simulations to determine numerical effects which are summarised in Appendix \ref{app:resolution}.
The lower resolution runs do not differ significantly from the high resolution runs, except that less substructure is resolved, so the high resolution runs are more suitable for exploring the change in satellite populations.

Gravity is modelled using a tree and particle mesh (tree-PM) algorithm using periodic boundary conditions with \textsc{Arepo} code \citep{Springel2010,Weinberger2020}.
A Friends-of-Friends (FoF) algorithm is used to identify haloes within the simulation \citep{Davis1985}.
In this algorithm, particles or substructures are linked when they lie within a distance of $b\left(V/N\right)^{1/3}$, with $b$ set to $0.2$, $V$ is the volume of the box, and $N$ is the number of particles.
Substructures are identified using \textsc{SubFind}, which groups particles that are gravitationally bound to each other \citep{Springel2001,Dolag2009}.
The most bound particle determines the position of each halo.
The main halo is identified as the most massive gravitationally bound object in the FoF group, and other objects are subhaloes.

\subsection{SIDM model}

\subsubsection{Dark matter models and cross sections}
We are interested in studying the phenomenology of DM models with significant up-scattering cross-sections.
Such interactions arise naturally in a model of inelastic DM with nearly degenerate ground ($\chi^1$) and excited ($\chi^2$) state particles that are separated by a mass-splitting $\delta$.
The interactions between the two states proceed through a mediator $\phi$ (for instance, a kinetically mixed dark photon) with a strength proportional to $\alpha$.
Our parameter space therefore consists of four quantities: $m_{\chi^1},\,\delta,\,m_\phi$ and $\alpha$. 

Once DM is produced in the early Universe, the following types of interactions can occur in a halo:
\begin{itemize}
    \item Elastic scattering between two ground (excited) state particles, $\chi^i + \chi^i \to \chi^i+ \chi^i$, where $i = 1(2)$ respectively.
    \item Elastic scattering between a ground and excited state particle, $\chi^1 + \chi^2 \to \chi^1 + \chi^2$. 
    \item Inelastic up-scattering of ground state particles, $\chi^1 + \chi^1 \to \chi^2 +\chi^2$.
    \item Inelastic down-scattering of excited state particles, $\chi^2 + \chi^2 \to \chi^1 +\chi^1$.
\end{itemize}
In contrast to elastic SIDM models and CDM where only (elastic) gravitational interactions are allowed, our model allows for interactions where the kinetic energy of the system is not conserved. Up-scattering reactions are endothermic and remove energy from the system. Consequently, they are kinematically allowed only when the total energy of the ground state particles is larger than the mass-splitting, resulting in a velocity threshold
\begin{equation}
    v > 2\sqrt{2\delta/m_{\chi^1}}
    \label{threshold}
\end{equation}
where $v$ is the relative velocity between the particles in the centre of mass frame. On the other hand, down-scattering reactions are exothermic and provide a "velocity kick" to each of the final state particles
 \begin{equation}
     v_\mathrm{kick} = \sqrt{2\delta/m_{\chi^1}} \ .
 \end{equation}
 
In this work, we fix the interaction strength to be $\alpha=0.17$, which can generate interaction cross sections that are large enough to affect small scale structure while not being so large as to force small simulation time steps due to short mean free times.
Furthermore, we will utilise the analytic cross-sections for inelastic scattering derived in \cite{Schutz2014} in the Born regime, i.e. when $\alpha m_{\chi^1}/m_\phi \ll 1$, which implies that $m_\phi \gtrsim m_{\chi^1}$ for our choice of coupling. We work in the frequently studied limit where $m_\phi \sim 3 m_{\chi^1}$, and in this heavy-mediator regime the second Born approximation used for elastic scattering between pairs of ground state particles (and excited state particles) employed in \cite{Schutz2014} is not applicable because it only includes a subset of the diagrams that contribute. We therefore supplement the calculation with expressions from \cite{Fitzpatrick2021} for elastic scattering between particles in the same state, which reduce to the expressions in \cite{Schutz2014} in the light-mediator limit modulo a correction factor of 1/2 to account for the indistinguishability of these particles. For off-diagonal elastic scattering between the ground and excited states, we use the results from the Born approximation for tree-level scattering derived in \cite{Feng:2009hw}; we have checked that the presence of a mass splitting between states does not affect $\sigma/m$ to leading order in $\delta/m_{\chi^1}$. Because the parametric scaling for the up-scattering cross section is $\sigma \sim \alpha^2 m_{\chi^1}^2/ m_\phi^4$, cross sections with SIDM implications correspond to DM and mediator masses in the $\sim$MeV range. Mass splittings that are comparable to the kinetic energy of a DM particle in a halo with $v\sim 10^{-3}$ will therefore be in the $\lesssim$~eV range in order to be above the velocity threshold in Eq.~\eqref{threshold}.

\begin{figure}
    \centering
    \includegraphics[width=\linewidth]{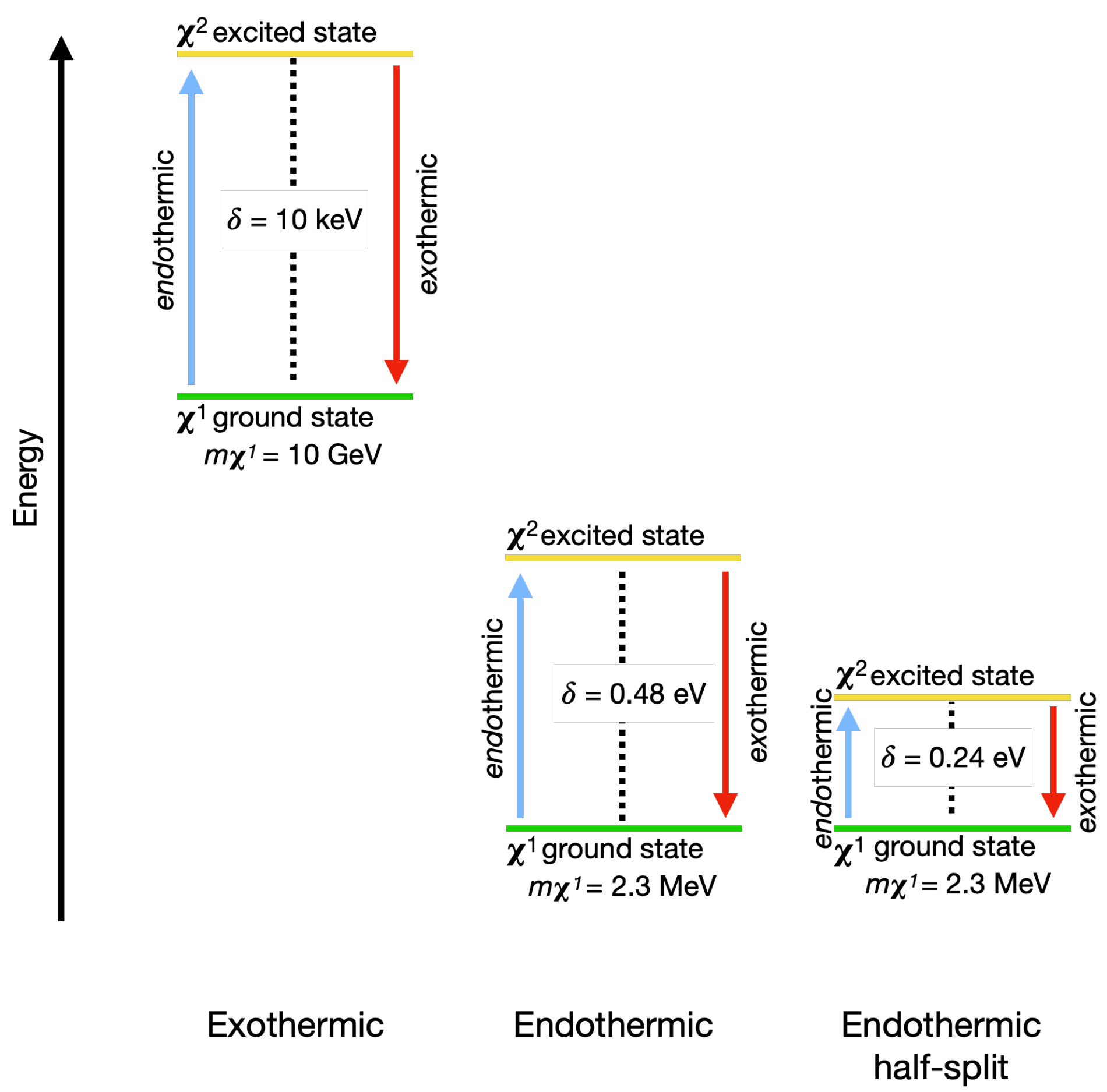}
    \vspace{-0.5cm} 
    \caption{A schematic showing the ground and excited states and transitions for each model.  Ground states are shown with the green horizontal lines and excited states are shown in yellow.  Endothermic reactions (blue, up arrow) transition from the ground to excited state and exothermic reactions (red, down arrow) transition from the excited to ground state.
 The Exothermic model (left) has a high ground state mass and a large mass splitting between the ground and excited states.  The Endothermic model has a lower ground state mass and mass splitting, and the Endothermic half-split model has the same ground state mass with half the mass splitting.}
    \vspace{-0.5cm} 
    \label{fig:reactions}
\end{figure}

\begin{table}
    \centering
    \begin{tabular}{c|cccc}
          & $m_{\chi^1}$ & $m_{\phi}$ & $\delta$ & $\alpha$ \\
          \hline
         Exothermic &  10 GeV &  30 MeV & 10 keV & 0.10  \\
         Endothermic & 2.3 MeV & 7.3 MeV & 0.48 eV & 0.17 \\
         Half-split & 2.3 MeV & 7.3 MeV & 0.24 eV & 0.17
    \end{tabular}
    \caption{Parameters for the Exothermic and Endothermic models. These are the DM ground state mass $m_{\chi}$, the dark force mediator mass $m_{\phi}$, the mass splitting between the ground and excited state $\delta$, and the coupling strength $\alpha$. In the half-split model, $\delta$ is halved compared to the default Endothermic model parameters.}
    \label{tab:parameters}
\end{table}

In order to distinguish the effects of these various elastic, exothermic, and endothermic processes, we divide our parameter space into five regions.
A schematic of the particle states and transitions is shown in Figure \ref{fig:reactions} and a summary of the chosen parameters is shown in Table \ref{tab:parameters}.
Using these parameters in the regimes and approximations as described above results in the set of elastic and inelastic cross sections shown in Figure \ref{fig:reactions}.

\begin{figure*}
    \centering
    \includegraphics[width=\linewidth]{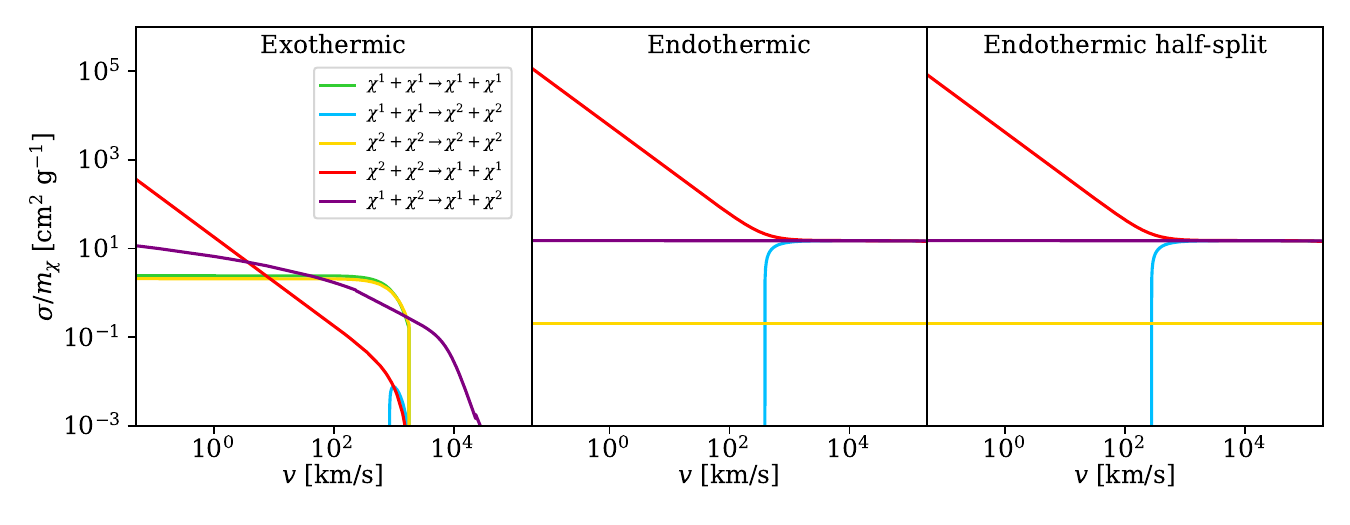}
    \caption{Scattering cross sections per mass ($\sigma/m$) as a function of relative velocity for DM particles. We show $\sigma/m$ for the exothermic model from \citet{Vogelsberger2019} (left), $\sigma/m$ for our endothermic model (middle) and $\sigma/m$ for the endothermic model with an excited state with half the mass splitting (right). Note that for our two new models (middle and right), the elastic scattering of particles in the same state (green and yellow lines; note that the green line in these plots is directly beneath the yellow line) is suppressed and the up-scattering cross section (blue line) is much greater than for the exothermic model at higher velocities. The half-split model has an up-scattering reaction that becomes significant at a lower relative velocity.}
    \label{fig:cross_sections}
\end{figure*}

\textit{Exothermic}: By considering a large enough mass-splitting, $\delta > v^2_{\chi^1}m_{\chi^1}/8$, it is possible to ensure that only elastic scattering and inelastic down-scattering processes are allowed in a halo, thus making the model effectively exothermic. Such a model has been previously studied in \cite{Vogelsberger2019} and acts as a useful benchmark for comparison with the case when significant up-scattering is allowed. The corresponding cross-sections are plotted as a function of velocity in the left panel of Fig. \ref{fig:cross_sections} for the benchmark values listed in Table \ref{tab:parameters}. Because down-scattering is the more relevant inelastic process, the initial conditions for the Exothermic simulation began with all particles in the excited state.  This differs from the other simulations, which all start in the ground state.  However, we do not alter the transfer function used to generate the initial conditions so the initial distribution and velocities of the particles are the same.  We use the comparison between this model and the endothermic models to emphasise the difference that occurs when the dark matter particles are dominated by different scattering reactions.  If the particles in this simulation were to start in the ground state, the difference between this model and the CDM model would be negligible since the particles would barely scatter at all and, since the up-scattering reaction is suppressed, would not experience any inelastic scattering.

\textit{Endothermic}: By lowering the mass-splitting compared to the DM mass, we end up in a scenario where the average DM velocity in a halo is large enough for ground state particles to up-scatter. Note that in this case, \textit{all} the processes listed above are allowed. We use the benchmark parameters 
in Table \ref{tab:parameters} and plot the corresponding cross-sections as a function of velocity in the middle panel of Fig. \ref{fig:cross_sections}. To maximise the effects of up-scattering in this simulation, we initialise all DM particles in the ground state. As discussed below, this initial condition is the most natural expectation given the DM production mechanism and early-Universe thermal history in the relevant parameter space.

\textit{Endothermic up-scattering}:
To further isolate the effects of up-scattering, we run a simulation using only the up-scattering $\sigma/m$ from the Endothermic simulation setting all other $\sigma/m$ to zero.
Thus, up-scattering can occur for particles with a relative velocity greater than 388.6 km s$^{-1}$  
and no other reactions are possible. This simulation is also initialised with all particles in the ground state.

\textit{Constant up-scattering}: For an even more extreme case of Endothermic up-scattering, we remove the velocity threshold for up-scattering and set $\sigma/m$ for all relative particle velocities to 15.12 cm$^2 \rm{g}^{-1}$. 
This is a toy model used to explore the case where there is no velocity threshold for up-scattering.  As in the Endothermic up-scattering model, all other cross sections for elastic and down-scattering are set to zero. We note that up-scattering decreases a particle's kinetic energy, so particles are only allowed to up-scatter as long as the final state kinetic energy in the simulation frame is greater than zero. Particles with very low velocities are therefore forbidden from up-scattering since the reaction would cause them to have a negative kinetic energy in the simulation frame. This differs from the velocity threshold appearing in the cross sections of \citet{Schutz2014}, which depends on the relative velocity of two particles, as required by Lorentz invariance. Instead, the condition imposed here depends on the total kinetic energy of each particle in the simulation frame and is enforced in order to avoid complex velocities in the simulation. This is not a physical scenario since it violates Lorentz invariance but allows us to explore the extreme cases of up-scattering on the halo. This simulation initialises all particles in the ground state.

\textit{Endothermic half-split}:
To study the effects of the velocity threshold in up-scattering, we decrease the mass splitting between the ground and excited state by half thereby decreasing the minimum velocity required for up-scattering. This comparison is useful for distinguishing between the effects of the scattering reactions themselves and those arising from the presence of a velocity threshold. The corresponding cross-sections are plotted in the right panel of Fig. \ref{fig:cross_sections} for the benchmark parameters listed in Table \ref{tab:parameters}.

\textit{CDM model}:
Additionally, we use the standard \textit{CDM model} which includes only gravitational forces for comparison.

We note that the abundance of ground and excited state particles in a halo depends on the cosmological history of the DM model under consideration, i.e., the initial production of $\chi^{1,2}$ from the SM heat bath and any subsequent inter-conversion, $\chi^1 \leftrightarrow \chi^2$. As such, there may be additional model-specific bounds on the parameter space. For the purposes of this paper, we assume that inelastic DM is produced via the freeze-in mechanism \citep{Hall:2009bx,An2020,Heeba:2023bik}, in contrast to thermal freeze-out (see for example \citealt{Baryakhtar2020,Fitzpatrick2021}) since there are strong constraints on $\sim$MeV scale thermalized dark sectors from BBN and the CMB \citep{2020JCAP...01..004S}. The freeze-in mechanism additionally ensures that our benchmark values for the parameters are not in conflict with any constraints on our particle physics model and reproduce the measured DM relic density \citep{Planck:2018_cosmo_params}. Specifically, for the freeze-in mechanism to achieve the observed relic abundance with our choice of dark coupling constant $\alpha = 0.17$, the kinetic mixing parameter for the dark photon mediator will be $\sim10^{-11}$, which is small enough to avoid constraints from excessive energy loss from Supernova~1987A from MeV-scale dark photons \citep{1611.03864} and the DM itself \citep{Chang:2018rso} while being large enough to avoid BBN constraints on MeV-scale dark photons \citep{Fradette:2014sza}. Moreover, DM made via freeze-in can have a parametrically smaller temperature at late times than DM of the same mass made by freeze-out due to the absence of a kinetic coupling between frozen-in DM and the SM bath. Such a kinetic coupling between the two sectors will cause the DM temperature to evolve with scale factor $a$ as $T_\chi \sim 1/a$, even when the DM is nonrelativistic, whereas if nonrelativistic DM is \emph{decoupled} from the SM its temperature will evolve as $T_\chi \sim 1/a^2$. By the time the Universe reaches a temperature of $\sim 1$~eV, near the scale that is relevant for recombination, frozen-in MeV-scale DM will have a temperature of order $10^{-6}$~eV. Therefore, DM with an eV-scale mass splitting will have its excited state severely depleted by $\chi^2 + \chi^2 \to \chi^1 +\chi^1$ due to Boltzmann suppression. This prevents annihilation to SM particles from occurring close to the time of recombination, since the annihilation can only proceed with a mix of ground and excited state particles. Therefore, this scenario is compatible with CMB constraints on annihilating DM \citep{Slatyer:2009yq,Slatyer:2015jla,Planck:2018_cosmo_params}. We note that other inelastic DM thermal histories beyond freeze-in, for instance ones involving resonant DM annihilation in the early Universe, may also lead to interesting SIDM behavior with varying amounts of DM in the ground and excited states~\citep{resonantIDM}.

\subsubsection{Implementation in the simulations}
\label{sec:methods_sidm}

To model the DM interactions, we use the same algorithm used in \citet{Vogelsberger2019}, which we summarise here. 
Though the physical model is based on fundamental DM particles, in practice, these particles are represented by macroscopic simulation particles with masses as given in Section \ref{sec:methods_sims}.
Although they represent many fundamental DM particles, each simulation particle exists in only one state (ground or excited) and represent a population of particles entirely in that state.
This population of fundamental particles similarly scatters together with the simulation particle.
The scattering frequency determined from $\sigma/m_\chi$ is therefore dependent on the simulation particle mass rather than the fundamental $m_\chi$ and these scattering events happen less frequently than would occur for individual particles.
Over the course of the simulation, the total amount of dark matter that scatters is consistent with if they had scattered as fundamental particles.
Only the zoom particles in the simulations experience scattering, and the background particles interact only gravitationally as in CDM.

The probability of a particle scattering depends on the scattering cross section and the nearby particles.
We identify the neighbour particles (between 10 and 38 nearest neighbours) and the neighbour particles contribute to the scattering probability according to a cubic spline kernel function.
The resulting probability of two particles scattering in a given reaction is then the product of the mass of the particle, the cross section of the reaction at the particles' relative velocity, and the distance the particles would move in one time step at this velocity and adjusted by the appropriate factor from the Kernel.
The probability of each scattering reaction or of not scattering sums to one for each particle.
A random number between 0 and 1 is drawn for each particle to determine whether it will scatter and, if so, what reaction will occur.
In the case that the particle scatters, a nearby particle in the appropriate state is chosen scatter with.
Scattering occurs isotropically, so the particle final velocities occur in any direction after scattering with equal probability.
In inelastic collisions, the kinetic energy of the scattering particles are updated to account for the mass difference between the states and is distributed evenly between the two particles.

The time steps in the simulation should be set such that particles do not scatter multiple times within each step to avoid the DM particles behaving more like a fluid than like largely collisionless particles.
This typically decreases the time steps compared to the limits from the gravitational interactions.
In the event that a particle attempts to scatter more than once in a single time step, the second scatter is prevented from occurring.
In testing our models, we ensure that the number rejected scattering events do not reach more than a few percent of the total scattering events for any given reaction.
This also informed our choices of parameters shown in Table \ref{tab:parameters} since we required our parameters to produce cross sections that emphasised the reactions of interest without becoming so large as to make the time steps prohibitively small.
We also checked that this corresponds with the overall energy injected into the particles to ensure that there is proper conservation during the simulation.
\section{Results}
\label{sec:results}

\begin{figure*}
    \includegraphics[width=\linewidth]{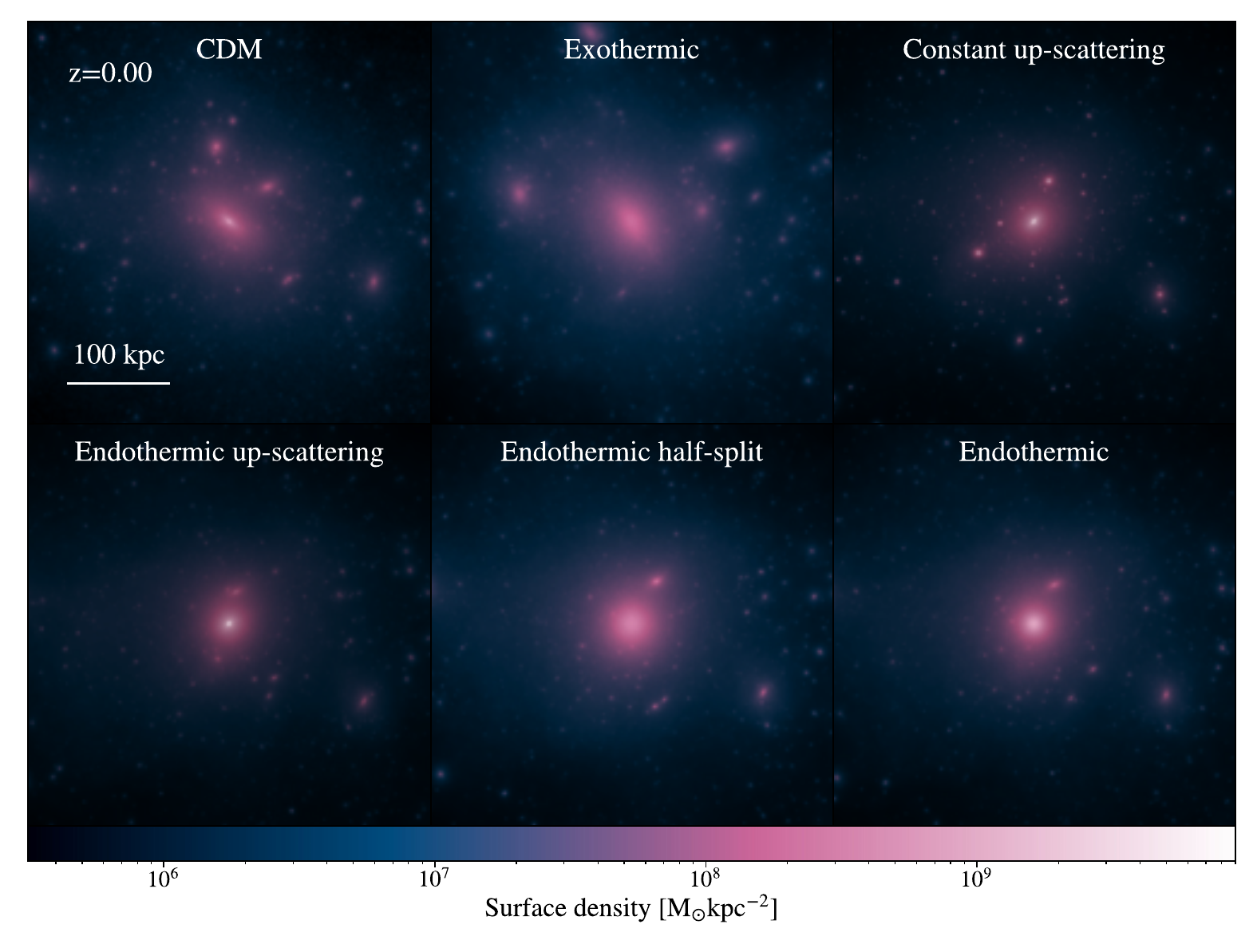}
    \vspace{-.5cm}
    \caption{DM surface density projections for the main halo in each run at $z=0$.  The Exothermic model does not reach densities as high as in the CDM case, especially in the centres of haloes in both the main halo and in subhaloes.  The Endothermic models all show very dense regions in the centres of their haloes since the up-scattering reaction causes particles to fall to smaller radii.  Several of these models also have suppressed subhalo formation due to the other reactions (elastic and down-scattering).  The Constant up-scattering model produces as many subhaloes as the CDM model, and these subhaloes also have higher central densities since particles of all velocities are able to up-scatter, while the Endothermic up-scattering does not see this effect in its subhaloes due to the velocity threshold for up-scattering.}
    \label{fig:projections}
\end{figure*}

We show in Figure \ref{fig:projections} the projections of the main halo and its satellites for each of our six models.
Notably, the models with significant up-scattering (Endothermic, Endothermic up-scattering, Constant up-scattering and Endothermic half-split) reach much higher densities in the inner regions of their main haloes.
There are also noticeably fewer subhaloes in the SIDSM models compared to the CDM model, except for the Constant up-scattering model where the subhaloes appear in similar numbers with higher densities.
The main halos with significant up-scattering also appear to be more spherical than the CDM and Exothermic model.
This may be due to endothermic scattering early in the halo's formation that causes particles to fall inwards and create a deeper potential well.
This would cause the halo to become more spherical even in the models where there is also exothermic scattering since the endothermic reactions dominate early in the halo's formation.
In this Section, we explore in more detail the properties of the main halo and subhaloes for our models and how they are affected by each DM model.
We investigate trends in the density between each model as well as the subhalo abundances, densities and velocity dispersions.
We relate these trends to the reactions that occur through time and show how the inner regions of the halo change with redshift.

\subsection{Main halo density}
\label{sec:results_halo}

\begin{figure}
    \vspace{-.5cm}
    \centering
    \includegraphics[width=\linewidth]{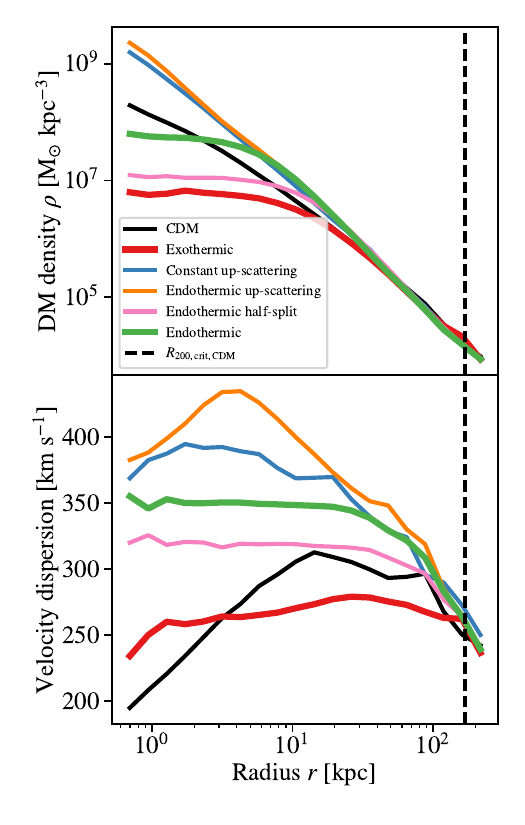}
    \caption{DM density (top) and average velocity dispersion of particles (bottom) as a function of distance from the main halo centre for the main halo for each model at $z=0$.
    The Exothermic model has a significantly decreased central density since elastic and down-scattering kick particles to the outer regions of the halo.  This corresponds to a low velocity dispersion for a wide range of radii and low overall ground state fraction.
    The Endothermic up-scattering and Constant up-scattering have high central densities with steep slopes since up-scattering causes particles to fall inwards.
    These models show notable peaks in their velocity dispersion profiles since few scatters are able to occur at small radii.
    At these small radii, a high percentage of particles quickly reach the excited state, which halts further scattering.
    The up-scattering increases the density in the Endothermic model, but the elastic and down-scattering also give this model a core.
    The Endothermic half-split model similarly has increased density but less so because the increased up-scattering also increases the down-scattering, which decreases density.  The velocity dispersions for these models are higher than the Exothermic model corresponding to higher ground state fractions.}
    \label{fig:halo_profiles}
\end{figure}

\begin{figure}
    \vspace{-.5cm}
    \centering
    \includegraphics[width=\linewidth]{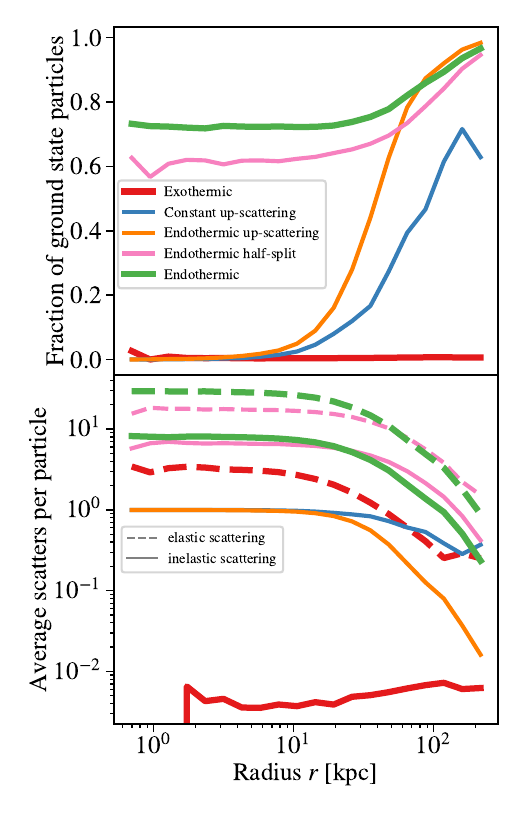}
    \caption{The fraction of particles in the ground state (top) and the average number of elastic (dashed) and inelastic (solid) collisions each particle has had (bottom) as a function of distance from the halo centre at $z=0$.  With all particles in the excited state in the Endothermic up-scattering and Constant up-scattering models, not much scattering occurs there, which causes the velocity dispersion to fall as shown in Figure \ref{fig:halo_profiles}.  At radii less than $r\approx10$ kpc, particles have scattered once in their history and are unable to scatter again.  The Endothermic model has a slightly higher average number of scatters per particle compared with the Endothermic half-split model, corresponding to a higher ground state fraction and increased velocity dispersion. The Exothermic model has low levels of scattering, so most particles are in the excited state where they started and the velocity dispersion is correspondingly low.  These three models have high levels of elastic scattering, which causes the haloes to be isothermal and flattens the velocity dispersion profiles in Figure \ref{fig:halo_profiles}.}
    \label{fig:sidm_profiles}
\end{figure}

Figure \ref{fig:halo_profiles} shows the total DM density and velocity dispersion as a function of distance from the main halo centre.
For each radius, the spherically averaged density is calculated using all DM particles within even logarithmically spaced spherical shells.
The velocity dispersion is calculated by averaging the difference in velocity in the frame of the halo from the average velocity in each shell.
The black line shows the density for the main halo using CDM, and the coloured lines show the density for each of our SIDM models.
Up-scattering is an endothermic reaction, so the resulting particles lose kinetic energy and decrease velocity.
This causes them to fall to a smaller orbit within the halo and increase the density at small radii.
Down-scattering will have the opposite effect, with particles gaining kinetic energy and getting kicked to larger radii, creating a core in the centre of the halo.

The Exothermic model, as demonstrated in \citet{Vogelsberger2019}, results in significantly lower densities and velocity dispersions than the CDM model.
The models with only up-scattering cross sections, the Endothermic up-scattering and Constant up-scattering, show increased central densities and increased velocity dispersions at intermediate radii.
With just the up-scattering reaction, the density profiles remain steep to small radii, while the addition of down-scattering and elastic scattering gives the density profile in the Endothermic model a core.
This also decreases the velocity dispersion at lower radii since not all particles are scattering and changing velocity.
The Endothermic model also has an increased density compared to the CDM model for much of the density profile, but the inner region is much more cored due to the elastic and down-scattering.
The Endothermic half-split model results in significantly lower densities in the inner regions of the halo compared to the Endothermic model.
The lower velocity threshold for up-scattering in this profile allows more of these reactions, which alone would cause a higher central density.
However, more particles moving to the excited state also allows much more down-scattering, which decreases the central density.
Thus, instead of the Endothermic half-split model resulting in a density profile that falls between the Endothermic and the up-scattering models, we find that the coring effect of the Endothermic model compared to the up-scattering models is enhanced in the Endothermic half-split model.

We show the velocity dispersion of the particles as a function of radius in the bottom panel of Figure \ref{fig:halo_profiles}.
Generally, up- and down-scattering will increase the velocity dispersion since these reactions change the velocity of particles.
If particles are moving at speeds similar to nearby particles and then scatter, this increases the dispersion since they are now moving at different speeds.
The Exothermic model has a lower velocity dispersion at all radii since most particles are in the excited state and not many down-scattering reactions occur.
The Endothermic and Endothermic half-split models have increased velocity dispersions compared to the Exothermic model.
This corresponds to more up- and down-scattering.
The Endothermic model has a higher dispersion than the Endothermic half-split since this model has more particles in the ground state.
This allows more particles to up-scatter and, since down-scattering is overall more likely to occur (the down-scattering reaction is more likely at lower relative velocities), more scattering occurs.
The up-scattering only models, Endothermic up-scattering and Constant up-scattering, peak in velocity dispersion at intermediate radii.

We  examine the fraction of ground state particles as a function of distance from the halo centre in the top panel of Figure \ref{fig:sidm_profiles}.
The bottom panel shows the average number of times a particle has scattered over the course of the simulation as a function of radius.
In general, down-scattering gives particles a velocity kick, which pushes them to higher radii.
This causes down-scattering to increase the ground state fraction at larger radii more than at lower radii.
Up-scattering causes particles to fall to lower radii, which decreases the ground state fraction at lower radii.

For the Exothermic model (red line), almost all particles remain in the excited state throughout the simulation, so the ground state fraction is low at all radii.
This model is dominated by down-scattering and elastic scattering, with elastic scattering somewhat more significant at typical halo velocities.
Since all particles start in the excited state for this model and elastic scattering is more likely than down-scattering, the majority of particles remain in the excited state.

The Endothermic up-scattering and Constant up-scattering models (orange and blue lines) start entirely in the ground state and allow for only up-scattering reactions.
As more scattering occurs, particles will excite without the possibility of returning to the ground state.
For the Constant up-scattering model, the scattering rate depends only on the density of ground state DM particles while the Endothermic up-scattering has the additional constraint that particles must be moving with relative velocities close to 300 km/s to scatter.
This leads to the lower ground state fraction at intermediate radii for the Constant up-scattering model compared to the Endothermic up-scattering model since the particles are dense enough but are not moving with high enough velocities for consistent up-scatting with the velocity constraint.
At low radii, most particles are in the excited state, so no scattering can occur.
This causes the decrease in the velocity dispersion near a radius $r\approx8$ kpc (Figure \ref{fig:halo_profiles}) similar to the CDM velocity dispersion peak at slightly larger radius $r\approx11$ kpc.
Without much scattering occurring, the velocity dispersion is dependent primarily on the density of the halo as in the CDM case.
This occurs at slightly larger radii for the Constant up-scattering model since the velocity threshold allows particles with lower velocity differences to scatter.

The Endothermic and Endothermic half-split models (green and pink lines) have significant up- and down-scattering as well as elastic scattering.
At higher velocities, $v>389$ km/s for the Endothermic model and $v>275$ km/s for the Endothermic half-split, the up-scattering and down-scattering reactions are equally likely (see Figure \ref{fig:cross_sections}).
We see the results of this as the ground state fraction levels off at smaller radii for these models.
The fraction for the Endothermic model is higher than 50\% since down-scattering is significantly more likely for low velocity particles.
This means that there is a population of particles that may down-scatter to the ground state but will not up-scatter to the excited state, increasing the overall ground state fraction.
The Endothermic half-split model levels off closer to 50\% but still well above since the velocity threshold for up-scattering is lower, so the population of particles that may down-scatter but not up-scatter is smaller.

\begin{table*}
    \centering
    \begin{tabular}{c|cccccc}
         Model & $M_{\rm200,mean}$ & $r_{\rm200,mean}$ & $V_{\rm max}$ & $R_{\rm max}$ & $N_{\rm sub}$ & ground state population \\
         & [$10^{12}\ \rm{M}_{\odot}$] & [kpc] & [km/s] & [kpc] & & [\%] ($r<r_{\rm200,mean}$) \\
         \hline
         CDM & 2.16 & 401.27 & 178.78 & 87.75 & 2484 & -- \\
         Exothermic & 2.04 & 393.79 & 166.93 & 198.13 & 2289 & 0.60 \\
         Endothermic & 2.13 & 399.33 & 222.38 & 23.04 & 1639 & 86.87 \\
         Endothermic up-scattering & 2.16 & 401.26 & 246.49 & 13.49 & 1689 & 72.03 \\
         Constant up-scattering & 2.16 & 401.10 & 222.08 & 11.28 & 1895 & 44.76 \\
         Endothermic half-split & 2.11 & 398.34 & 199.69 & 45.86 & 1544 & 81.89 \\
    \end{tabular}
    \caption{Basic properties of the main halo in each model at $z=0$.  The mass $M_{\rm200,mean}$ and radius $r_{\rm200,mean}$ within which the enclosed overdensity is 200 times the mean density of the universe, the maximum circular velocity $V_{\rm max}$ reached at radius $R_{\rm max}$, the total number of resolved subhaloes $N_{\rm sub}$ within $r_{\rm200,mean}$ and the percent of ground state DM particles within $r_{\rm200,mean}$.}
    \label{tab:halo_properties}
\end{table*}

Table \ref{tab:halo_properties} shows some properties of the main halo at $z=0$.
We show the mass $M_{\rm200,mean}$, radius $R_{\rm200,mean}$, maximum circular velocity $V_{\rm max}$ and the radius at which it occurs $R_{\rm max}$, the number of subhaloes and the percent of particles within $R_{\rm200,mean}$ that are in the ground state.
The Endothermic up-scattering and Constant up-scattering models produce main haloes with the same mass as the CDM model while the others (Exothermic, Endothermic and Endothermic half-split) are slightly less massive.
Down-scattering increases the kinetic energy of particles, which kicks them to larger orbital radii and potentially out of the halo, leading to a decreased mass for the Exothermic model.
In the Endothermic and Endothermic half-split models, the up-scattering counteracts this effect to some extent since it causes particles to fall in to the halo.
However, falling in to the halo increases the density, leading to more scattering events, so an up-scattered particle is likely to down-scatter later.
Down-scattering, on the other hand, kicks particles to less dense regions and possibly out of the halo, where they are less likely to fall back in.
Thus, the combined effect of these reactions is a decrease in the mass of the halo but a smaller decrease than in the Exothermic model.
The models with only up-scattering do not result in mass loss since there is no down-scattering to kick particles out of the halo.
There is also no mass increase in these models since it does not make particles more likely to accrete onto the halo but rather makes the halo more concentrated.
For these models, the maximum circular velocity occurs at a smaller radius and reaches a higher value.

\subsection{Substructure}
\label{sec:results_substructure}

\begin{figure}
    \centering
    \includegraphics[width=\linewidth]{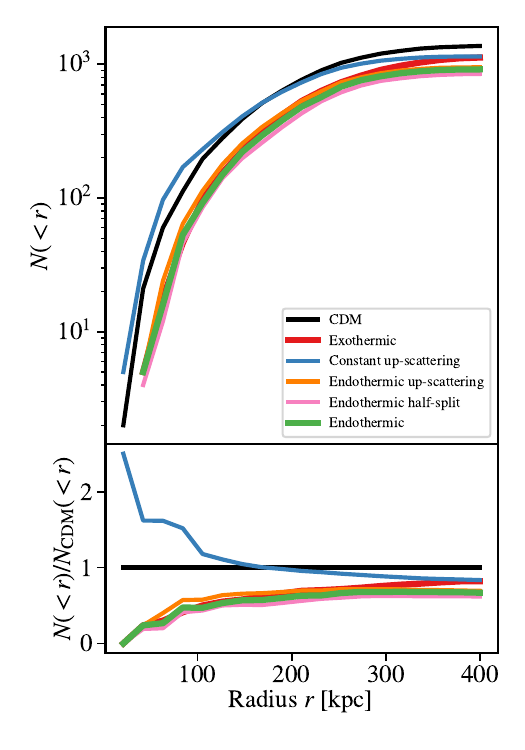}
    \caption{The number of subhaloes above $10^7\ \rm{M}_{\odot}$ within a given radius (top) and the same quantity divided by the corresponding value for the CDM model.  The Constant up-scattering model has a similar number of subhaloes as the CDM model with more subhaloes near the centre and fewer at larger radius since the main halo and structure is more condensed, while the other models have fewer subhales especially for $r<150$ kpc.  The Exothermic model converges with the CDM model at $r>200$ kpc, indicating that it has many subhaloes at large radius.}
    \label{fig:subhalo_number_profile}
\end{figure}

\begin{figure}
    \centering
    \includegraphics[width=\linewidth]{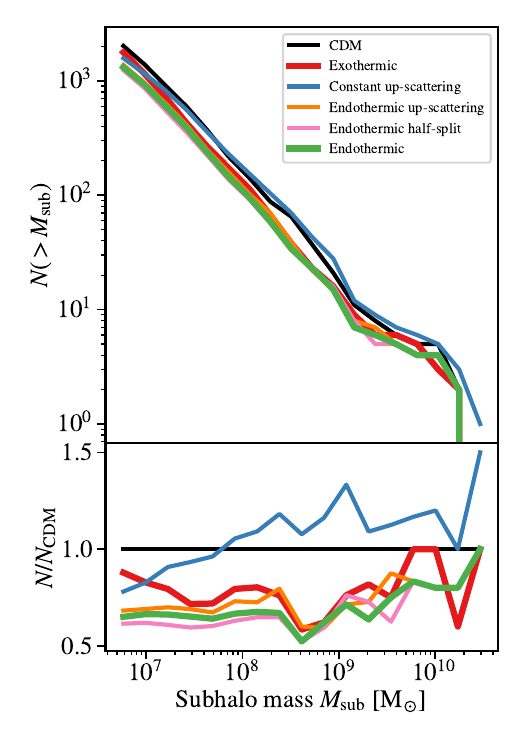}
    \caption{The number of subhaloes above a given mass for each model (top) and the number of subaloes above a given mass divided by the corresponding value for the CDM model.  The only model that does not suppress substructure is the Constant up-scattering model.  The scattering and altered density distributions in the other models result in a lower number of subhaloes.}
    \label{fig:subhalo_mf}
\end{figure}

The total number of subhaloes within $R_{\rm200,mean}$ of the main halo for each model is shown in the sixth column of Table \ref{tab:halo_properties}.
In Figure \ref{fig:subhalo_number_profile}, we show the number of subhaloes with bound mass $M>10^7\ \rm{M}_{\odot}$ within a given radius, and we show the number of subhaloes associated with the main halo above a given mass in Figure \ref{fig:subhalo_mf}.

The models have similar numbers of large subhaloes ($M_{\rm sub}\approx10^{10} \rm{M}_{\odot}$), and all models except the Constant up-scattering have slightly fewer lower mass subhaloes than the CDM model.
We expect down-scattering to decrease the amount of substructure since this increases the kinetic energy of particles and makes it more difficult to bind them in small haloes through gravity.
\citet{Vogelsberger2012b} also showed that elastic scattering suppresses the substructure in haloes for $\sigma/m$ of $10\ {\rm cm}^2\rm{g}^{-1}$.
This is roughly the same as the elastic scattering $\chi^1+\chi^2\rightarrow\chi^1+\chi^2$ in the Endothermic model ($14.89\ {\rm cm}^2\rm{g}^{-1}$).
The ground and excited state scattering cross sections per mass are suppressed at $0.21\ {\rm cm}^2\rm{g}^{-1}$.
Thus, we expect the elastic scattering from the off-diagonal reactions to have a significant impact on the substructure of these haloes in addition to the effects from inelastic scattering.
The Exothermic model, as expected for a model with significant elastic and down-scattering, has a smaller number of subhaloes than the CDM model.
This is consistent with the findings in \citet{Vogelsberger2019}.

In the Endothermic model, although there is significant up-scattering, there is also significant elastic scattering and down-scattering especially at lower velocities typical in subhaloes.
By redshift $z=0$, there is a roughly equal amount of up-scattering and down-scattering, and most reactions are the elastic scattering between ground and excited state particles in the main halo.
Like the Exothermic model, this then decreases the overall subhalo abundance.
Similar effects occur in the Endothermic half-split model since the velocity threshold is not different enough to substantially alter subhalo behaviour.

Although the Endothermic up-scattering model, like the Constant up-scattering, consists only of up-scattering reactions, it also suppresses the lower mass end of the subhalo mass function in contrast to the Constant up-scattering results.
The model behaves consistently with the Constant up-scattering model in the density profile of the main halo (top panel of Figure \ref{fig:halo_profiles}) and consistent with expectations that up-scattering increases the density of the main halo by reducing the kinetic energy of DM particles.

\begin{figure}
    \centering
    \includegraphics[width=\linewidth]{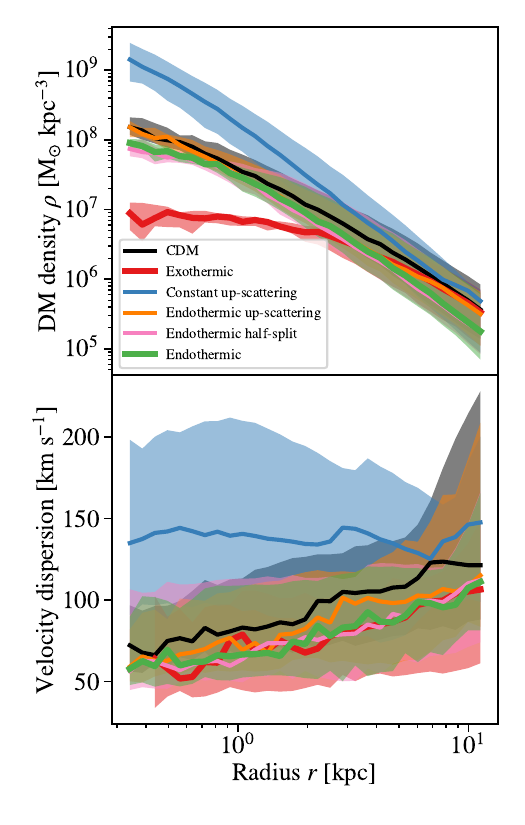}
    \caption{The density (top) and velocity dispersion (bottom) as a function of radius for the 10 most massive subhaloes.  The solid lines show the median value for each model and the shaded region shows the $16^{\rm th}$ to $84^{\rm th}$ percentiles.  The Exothermic and Constant up-scattering, which allow scattering at low velocities, show significant differences in the density profiles at low radii from the CDM case similar to the main haloes in the top panel of Figure \ref{fig:halo_profiles}.  The Endothermic and Endothermic up-scattering models show little difference relative to the CDM density profile since the relative velocities are not high enough to reach the velocity threshold for up-scattering. 
    The Endothermic and Endothermic half-split model decrease in density compared to CDM due to stronger tidal stripping.
    The velocity dispersion is highest in the outer regions of the subhaloes as the particles mix with particles from the main halo.  All models show similar trends, although the Constant up-scattering subhaloes have a slightly higher velocity dispersion at all radii.}
    \label{fig:subhalo_profiles}
\end{figure}

To explore this phenomenon in more detail, we show in Figure \ref{fig:subhalo_profiles} the density and velocity dispersion as a function of radius for the ten most massive subhaloes (top and bottom panels respectively).
We calculate the velocity dispersion for all particles associated with the subhalo, including main halo particles, so we reflect the scattering that occurs within this region.
The solid lines show the median value at each radius and the shaded regions span the $16^{\rm th}$ and $84^{\rm th}$ percentiles.
The Constant up-scattering model has significantly higher densities at lower radii than the CDM model similar to the main halo density profile.
Since there is no velocity threshold for up-scattering in this model, scattering can occur even in these smaller gravitational wells.
The velocity dispersion is also slightly higher at all radii in this model from the larger densities.
The subhaloes produced by the Endothermic and Endothermic up-scattering models are similar to those in the CDM model.
With the velocity threshold required for up-scattering, particles in these subhaloes are much less likely to scatter, and their behaviour is therefore similar to standard CDM.
Although the Endothermic up-scattering and Constant up-scattering produce similar main halo density profiles, they differ in the subhalo density profiles due to the velocity threshold for up-scattering in the Endothermic up-scattering model.
Even the Endothermic half-split model, which has a lower velocity threshold than the Endothermic model, shows little difference from the CDM model in the subhaloes.
The difference in density profiles between the main halo can therefore probe the velocity dependence of the scattering cross sections.

Dwarf haloes on circular orbits experience a tidal force that depends on the mass of the subhalo, the main halo mass enclosed by the subhalo's orbit and the orbital radius.
Tidal forces disrupt the mass within a subhalo and strip away the outer, less bound material.
This causes a drop in the density profile of the halo that is characterised by the tidal radius.
Higher tidal effects and stripping result in a smaller tidal radius.
As described in \citet{King1962} and \citet{Pace2022}, the tidal radius $r_t$ is given by:
\begin{equation}
    r_t = r \left(\frac{m_{\rm sub}}{2M_{\rm main}\left(<r\right)}\right)^{\frac{1}{3}}
    \label{eq:tidal_radius}
\end{equation}
where $m_{\rm sub}$ is the mass of the subhalo, $M_{\rm main}$ is the mass of the main halo and $r$ is the distance of the subhalo from the centre of the main halo.
Although this assumes a circular orbit, similar relative effects are found in elliptical orbits with increased stripping timescales \citep{Errani2021}.

The inner region of the main halo for up-scattering models is much more dense than in the CDM model, leading to a larger enclosed main halo mass at a given radius, even large radii, compared to the CDM model.
In the Endothermic up-scattering model, the subhaloes are similar to the CDM models due to the velocity threshold for up-scattering and are thus much more susceptible to tidal stripping.
This leads to a general decrease in subhalo masses.
With the innermost density roughly ten times larger in the Endothermic up-scattering case, we would expect this effect to alter the subhalo mass function by not more than a factor of $\sim2$, peaking in the inner region of the halo.
We see in the  bottom panel of Figure \ref{fig:subhalo_mf} that the abundance of substructure in this model is decreased by between $\sim0.6-0.8$.
This reaches close to the predicted factor of 2 and is lower since the relative density decreases farther out in the halo.
We can see from the bottom panel of Figure \ref{fig:subhalo_number_profile} that the subhaloes are more suppressed at smaller radius, where the relative density is higher.

The Constant up-scattering model, however, has much more dense subhaloes, making them less susceptible to this tidal effect.
\citet{Errani2020} showed that cuspier DM haloes are much more difficult to tidally disrupt compared to cored subhaloes.
These subhaloes are therefore not suppressed compared to the CDM case.
The inner density of the main halo is roughly ten times larger in the up-scattering models than in the CDM model (top panel of Figure \ref{fig:halo_profiles}).
The inner density of the Constant up-scattering subhaloes are also roughly ten times larger than the CDM subhaloes (top panel of Figure \ref{fig:subhalo_profiles}), cancelling out the tidal effects due to the increased main halo mass.

The Exothermic model produces subhaloes with similar properties to its main halo; the central densities are much lower than the CDM model.
Down-scattering occurs for relative velocities less than several hundred km/s and increases at lower velocities.
Elastic scattering is also significant in these conditions.
These reactions result in similar coring as in the main halo.
\subsection{Dark matter particle evolution}
\label{sec:results_particles}

\begin{figure*}
    \vspace{-.4cm}
    \includegraphics[width=\linewidth]{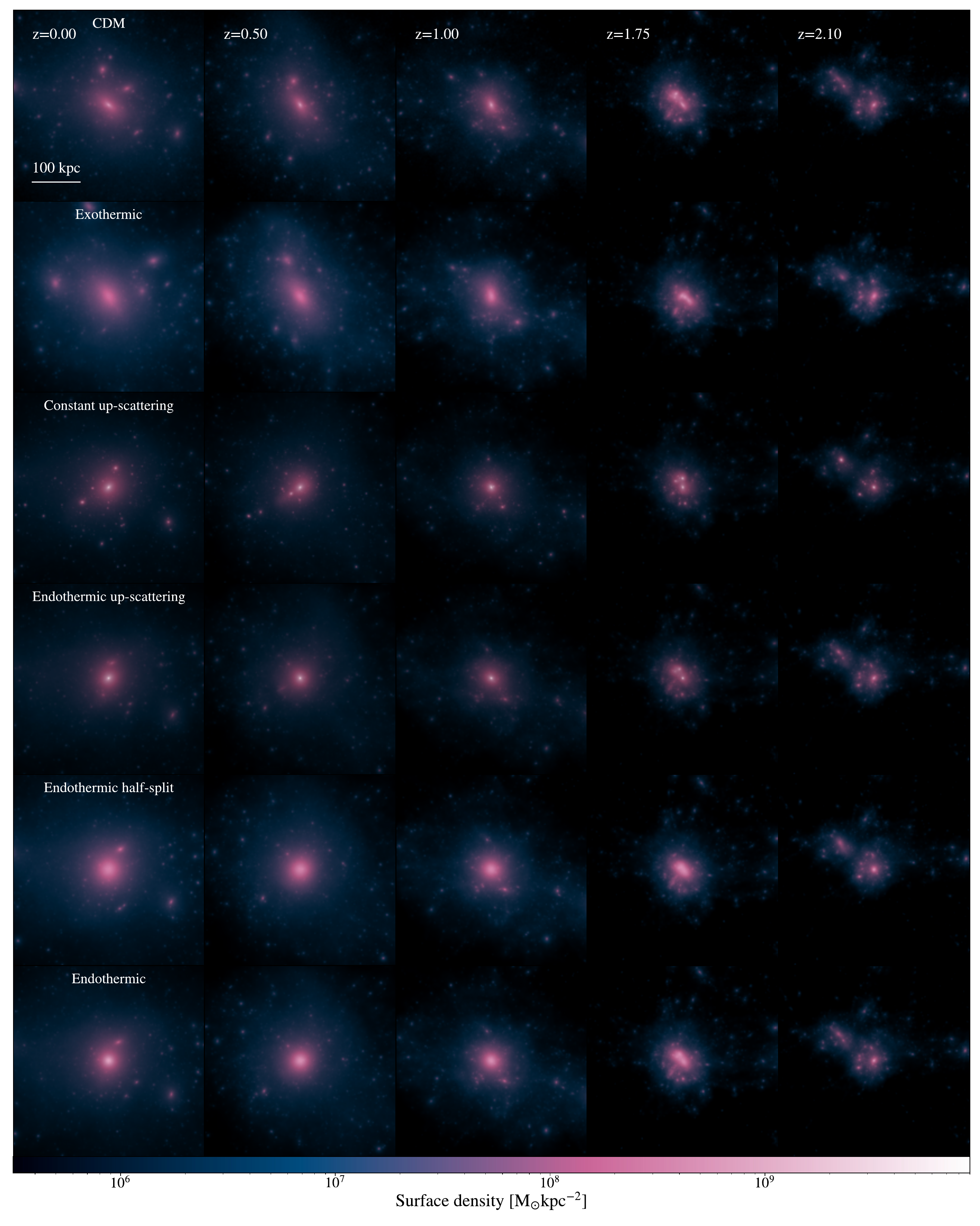}
    \vspace{-.5cm}
    \caption{DM surface density projections for the main halo in each model as in Figure \ref{fig:projections} at several redshifts between $z=0$ and $z=2.1$.  This redshift range shows the period of time in which the amount of scattering increases with a merger at $z\approx2$. Each model is shown on a given row with redshift increasing to the right in physical coordinates. The halo increases in size as it reaches $z=0$. The Exothermic model is less dense and the Constant up-scattering and Endothermic up-scattering are more dense for all redshifts. The Endothermic half-split is more dense at near $z=1$ and decreases in central density as down-scattering becomes more important, while the Endothermic model maintains a more consistent density through this time period.}
    \label{fig:projections_evolution}
\end{figure*}

To fully understand structure at $z=0$, we examine the evolution of the DM particles for the duration of the simulation.
In Figure \ref{fig:projections_evolution}, we show a visual of the main halo in each model in physical coordinates at representative redshifts between $z=0$ to $z=2.1$ (in each column).
Near $z\approx2$, the halo merges, which increases the density and velocity dispersion of the halo and allows more particles to up-scatter.
In this Section, we study the DM reactions that occur over time and connect this to the resulting halo structure.

\begin{figure*}
    \hspace{-.03\linewidth}
    \includegraphics[width=.9\linewidth]{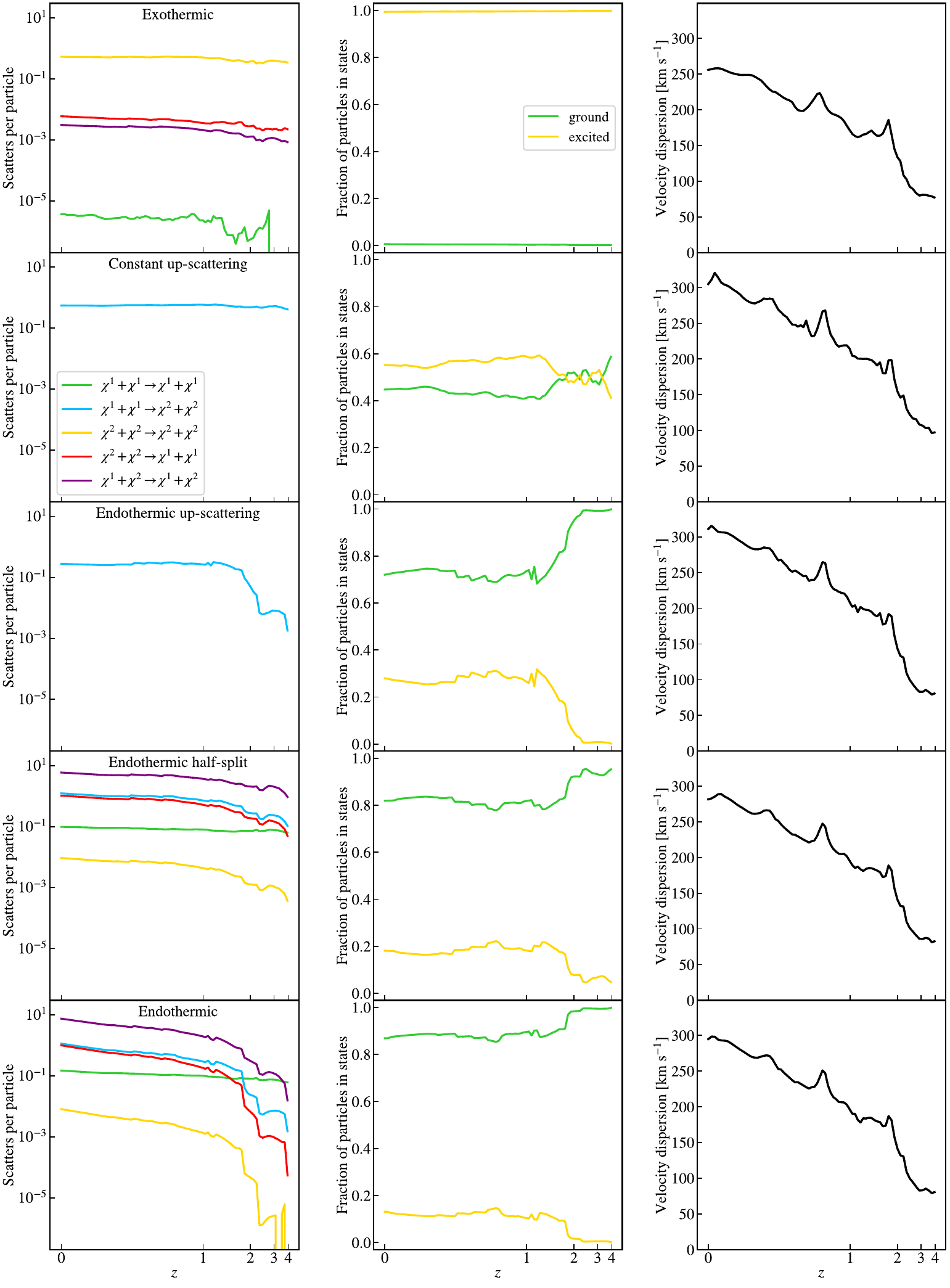}
    \caption[angle=90]{The evolution of DM particle properties for the main haloes in the SIDM runs.  We show on the left the total number of each reactions that occur, in the middle the fraction of ground and excited state particles and the average velocity dispersion for ground and excited state particles on the right.  The number of scatters increases with time as the halo collapses and the density and velocity dispersion increases.  The ground and excited state fractions change when scattering occurs frequently enough to have a significant effect.  There is a major merger at $z\approx2$ that triggers a sudden increase in scattering and a corresponding jump in the fraction of particles in each state.}
    \label{fig:particle_evolution}
\end{figure*}

\begin{figure}
    \center
    \includegraphics[width=\linewidth]{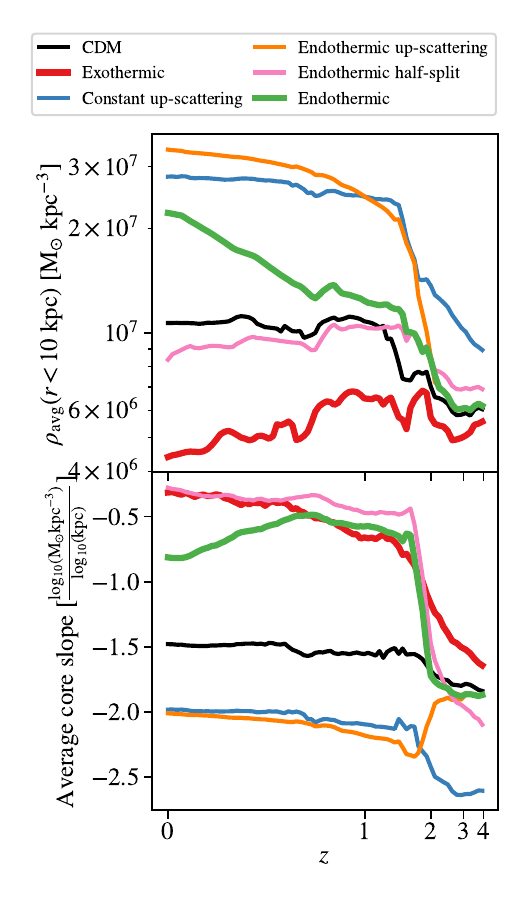}
    \caption{The average density (top) and density profile slope (bottom) within 10 kpc of each model's main halo centre as a function of redshift.  The Exothermic model is consistently lower than the CDM model due to elastic and down-scattering and develops a core with a slope near zero early on.  The Endothermic and Endothermic half-split models are similar to the CDM model until scattering becomes significant.  The Endothermic half-split model has a core similar to the Exothermic model due to significant elastic and down-scattering caused by more particles in the excited state.  Elastic and down-scattering decrease the core density while up-scattering increases it.  The Endothermic up-scattering and Constant up-scattering models, which allow only up-scattering and no elastic or down-scattering, are consistently more dense than the CDM model once scattering is significant.  These models also have steeper slopes near -2 compared to -1.5 for CDM.  The Constant up-scattering model has a steeper slope earlier on ($z=2-4$) compared to the Endothermic up-scattering due to the velocity threshold.}
    \label{fig:core}
\end{figure}

In Figure \ref{fig:particle_evolution}, we show the total number of each reaction per particle (left), particle states (middle) and particle velocity dispersion (right) through time.
Each model, except for the CDM model which has no scattering, is shown for a given row.
Figure \ref{fig:core} shows the average density and density profile slope of the main halo within 10 kpc.
This radius was chosen to correspond roughly to where the density profiles separate in Figure \ref{fig:halo_profiles}.
We use $R_{\rm half}$ to avoid pseudoevolution, which causes $R_{\rm200,mean}$ to expand with the expansion of the universe in addition to any changes within the halo.

To calculate the total number of reaction values, we find all particles in the main halo and sum the total number of scatters for the duration of the simulation for each reaction across all particles, then divide by the total number of particles.
The fraction of particles in each state is similarly the fraction of main halo particles in the ground or excited state.
To find the velocity dispersion, we take the velocity of all particles in the main halo, subtract the velocity of the halo centre, then we average the magnitude of each particle's velocity difference from the average velocity.

The dominant reaction in the Exothermic model is elastic scattering of excited state particles.
As shown in Figure \ref{fig:cross_sections}, elastic scattering dominates for velocities between tens and a few hundred km/s while down-scattering dominates for lower velocities.
Since particles start in the excited state, there is significantly more excited state elastic scattering (yellow line) than ground state elastic scattering (green line).
Although down-scattering is significant, it is still less likely at relative speeds of a few hundred km/s.
Thus, most particles remain in the excited state for the duration of the simulation.
The elastic scattering between ground and excited state particles roughly follows the amount of down-scattering through time.
Since the number of excited state particles is much greater than the number of ground state particles, the $\chi^1+\chi^2\rightarrow\chi^1+\chi^2$ reaction is driven by the abundance of ground state particles.
When particles down-scatter to produce ground state particles, these particles will be surrounded by excited state particles and are therefore likely to scatter with nearby excited state particles.
However, they are not likely to be near other ground state particles so there is little ground state elastic scattering or up-scattering, which is additionally suppressed due to a very small cross section.

Elastic scattering will create pressure in the DM halo and down-scattering will increase the kinetic energy of particles.
Both of these effects push particles to larger radii in the halo, creating more of a core and a lower central density.
This is evident in the top panel of Figure \ref{fig:core}, where the core density for the Exothermic model is consistently lower than the core density for the standard CDM model.

With only up-scattering, as shown in the second and third rows, the particles reach a high fraction of excited state particles.
In the Endothermic up-scattering, scattering starts only when particles reach a high enough velocity threshold and then amount of up-scattering is limited by the density of ground state particles.
This leads to a levelling off of the excited state fraction with time once the majority of particles at smaller radii are excited, as shown in the top panel of Figure \ref{fig:sidm_profiles}.
The Constant up-scattering model follows a similar trend except that particles do not need to reach a velocity threshold to up-scatter.
This means that the amount of up-scattering depends on the density of particles, so up-scattering begins earlier and the excited state fraction reaches a higher value and the increase in fraction occurs at a higher redshift ($z\sim4$ compared to $z\sim2$).
The particles in the Constant up-scattering model scatter even before they reach high velocities and begin siginificant scattering at moderate density at redshift $z\approx8$.
Near $z=2$, a merger occurs with the main halo.
This provides an influx of ground state particles that are a part of the merging halo, so the overall ground state fraction increases.
These new particles then provide a new population that is able to up-scatter, so the ground state fraction decreases again.
In the Constant up-scattering case, where the ground and excited state fractions are close to 0.5 early on, it is easy to see the changes in these fractions as particles fall into the main halo of the course of the merger.
Both the Endothermic up-scattering and Constant up-scattering models have central densities consistently higher than the CDM model (top panel of Figure \ref{fig:core}) once the excited state fraction becomes significant.

The Endothermic half-split model reaches a higher excited state fraction than the Endothermic model.
Because the velocity threshold is lower, more particles are able to scatter throughout the halo, as shown by the lower ground state fraction at all radii in the top panel of Figure \ref{fig:sidm_profiles}.
Particles also reach velocities that allow them to up-scatter at higher redshift compared to the Endothermic model.
This manifests with a noticeable increase in excited state particles before $z=2$, while the excited state fraction for the Endothermic model is flat at these times.
For much of the halo's evolution, the core density is similar to the CDM and Endothermic haloes.
At $z\sim2$, however, there's a distinct drop in the core density corresponding to the increase in scattering.
The Endothermic half-split model decreases more than the Endothermic model due to the somewhat higher amounts of elastic and down-scattering.

The Endothermic model, like the Endothermic half-split model, is initially dominated by elastic ground state scattering.
Although the cross section for this reaction is very small, all particles are in the ground state and this is the only reaction that is able to occur.
As the halo collapses and particles increase their speed, as shown with the average velocity dispersion of particles in the right panel of Figure \ref{fig:particle_evolution}, up-scattering becomes much more significant at redshift $z\sim2$.
This corresponds to a sudden increase in the fraction of excited state particles shown in the middle panel.
This also corresponds to an increase in elastic scattering.
Although the elastic (between ground and excited state particles), up- and down-scattering $\sigma/m$ are similar for relative velocities greater than $v\approx400$ km s$^{-1}$, up-scattering is not possible and the down-scattering cross section increases below this velocity while the $\chi^1+\chi^2\rightarrow\chi^1+\chi^2$ cross section remains constant.
Once a particle up-scatters, it is likely to elastically scatter with a nearby ground state particle.
This leads to more significant elastic scattering than up-scattering although the trends are similar.
Since down-scattering is significant at all velocities, especially lower velocities, the primary constraint on the amount of down-scattering is the abundance of excited state particles.
At higher velocities, the up-scattering and down-scattering $\sigma/m$ are equal.
Since the majority of particles in this model reside in the dense core where the velocities are particularly high, 
enough particles reach the excited state that down-scattering is equally as likely as up-scattering, and the reactions reach equilibrium.

\section{Conclusions}
\label{sec:conclusions}

We simulate a suite of Milky Way-\textit{like} haloes with varying DM models.
The alternative DM models are based on a two-state SIDM particle that can scatter elastically and inelastically.
We compare the effects of five models.
We use standard CDM as our fiducial model.
The Exothermic model has a significant down-scattering reaction cross section in addition to elastic scattering.
The Endothermic model has significant up-scattering at velocities $v>389$ km s$^{-1}$ as well as down-scattering and elastic scattering between ground and excited state particles.
The Endothermic up-scattering model sets all cross sections per mass from the Endothermic model except for up-scattering to zero.
The Constant up-scattering model has all cross sections per mass set to zero and the up-scattering cross section is set to the maximum value from the Endothermic up-scattering.
Finally, the Endothermic half-split model is similar to the Endothermic model but with a lower velocity threshold allowing up-scattering reactions.

We summarise our findings here:
\begin{itemize}
    \item Elastic and down-scattering create a core in the main halo, and up-scattering increases creating a main halo that is even cuspier than CDM.
    We show the density profiles of the main halo for each model in Figure \ref{fig:halo_profiles}.
    The Exothermic model has predominately elastic and down-scattering, and this model has the lowest and most cored central density.
    The Endothermic up-scattering and Constant up-scattering are only able to up-scatter, and they have the steepest and highest central densities.
    The Endothermic and Endothermic half-split models allow elastic and down-scattering in addition to up-scattering with similar cross sections.
    This results in a higher central density compared to the Exothermic model from the up-scattering reactions that is cored from the elastic and down-scattering.
    The Endothermic model has a higher central density than the CDM model while the Endothermic half-split model has a lower density since the lower velocity threshold for up-scattering leads to more particles in the excited state which then down-scatter or scatter elastically
    
    \item Increasing the amount of up-scattering decreases the equilibrium ground state fraction.
    In Figure \ref{fig:sidm_profiles}, we show the ground state fraction and average scatters per particle as a function of radius.
    In the inner regions of the halo, where the density and velocity dispersion is highest, the up- and down-scattering reactions are in equilibrium.
    The Endothermic half-split model has a lower velocity threshold for up-scattering so more particles are able to up-scatter, and this corresponds to a lower fraction of ground state particles at lower radii.
    The Endothermic up-scattering and Constant up-scattering each have a ground state fraction near zero at low radii.
    In the denser regions of the haloes, most particles have scattered with another nearby particle to leave the ground state and are then unable to down-scatter.
    The ground state fraction is lower at larger radii in The Constant up-scattering model compared to the Endothermic up-scattering model since more particles are able to up-scatter with no velocity threshold.
    
    \item Up-scattering condenses DM into more concentrated regions, while elastic scattering and down-scattering makes it more difficult for small structures to form through gravitational attraction.
    In models with both up- and down-scattering, we find that substructure is suppressed compared to the CDM case.
    With only up-scattering, the introduction of a velocity threshold also causes substructure to be suppressed even without elastic and down-scattering, which is primarily due to tidal effects.
    We show the number of subhaloes above a given mass for the main halo in each model in Figure \ref{fig:subhalo_mf}.
    The increased central density of the endothermic SIDM models increases tidal effects on subhaloes.
    Because of the velocity threshold in these models, the densities of subhaloes are not increased (Figure \ref{fig:subhalo_profiles}) and are more easily disrupted by the main halo which is denser than in the CDM case.
    
    \item Noticeable scattering begins at earlier times in models with lower velocity thresholds.
    The amount of elastic and down-scattering correlates with the formation of a core while up-scattering correlates with an increase in central density.
    We show in Figure \ref{fig:particle_evolution} the scatters for each reaction, the ground and excited state fractions, and the ground and excited state velocity dispersions as a function of redshift.
    Figure \ref{fig:core} shows the central density and density profile slope as a function of redshift for the main halo in each model.
    Since the endothermic models begin in the ground state, initial scattering is limited by the required kinematics for the up-scattering reaction.
    Once particles up-scatter, elastic scattering betweeen excited and ground state particles as well as down-scattering quickly follows.
    
\end{itemize}

In this work, we considered DM-only effects and leave explorations of the interplay between baryons and SIDM to future work.
SIDM can affect the behavior of baryons inside galaxies in a highly nontrivial way.
Even with only gravitational interaction between baryons and DM, the distribution and velocity of DM particles will affect the distribution and densities of stars and gas.
SIDM with constant cross sections per mass may, for example, increase the star formation in haloes compared to CDM haloes \citep{Sameie2021}, although this depends on the SIDM model.
In addition, baryons may alter halo behavior and observational signatures of SIDM compared to DM-only predictions \citep{Rose2022}.
\citet{Fry2015} found that baryons make the central densities of haloes indistinguishable from CDM for constant cross section per mass and \citet{Kamada2017} found that the addition of baryons contributed to the diversity of halo density profiles, although others have found that SIDM is more robust to changes in baryonic physics than CDM \citep{Robles2017}. 
Exploring only the dark component of haloes in this work isolates the effect of the DM model and helps disentangle the impact of these effects from baryonic physics.

Although up-scattering alone is not expected to mitigate small-scale problems with the CDM paradigm, the interplay between up-scattering with elastic and down-scattering is more difficult to predict.
Using realistic models with large cross sections, we show that DM particles with significant endothermic reactions have distinct effects on the halo and should therefore be observationally distinct from CDM, elastic scattering, or exothermic models.
However, full hydrodynamic simulations would be needed to better predict these signatures.

\section*{Acknowledgements}
It is a pleasure to thank Asher Berlin for useful conversations relevant to this work.
MV acknowledges support through NASA ATP 19-ATP19-0019, 19-ATP19-0020, 19-ATP19-0167, and NSF grants AST-1814053, AST-1814259, AST-1909831, AST-2007355 and AST-2107724.
SH acknowledges fellowship funding from the McGill Space Institute. SH and KS acknowledge support from a Natural Sciences and Engineering Research Council of Canada (NSERC) Subatomic Physics Discovery Grant.
PT acknowledges support from NSF grant AST-2008490.
MM acknowledges support through DOE EPSCOR DE-SC0019474 and NSF PHY-2010109.
TRS’ work is supported by the U.S. Department of Energy, Office of Science, Office of High Energy Physics of U.S. Department of
Energy under grant Contract Number DE-SC0012567; by a grant from the Simons Foundation (Grant Number 929255, T.R.S); and by the National Science Foundation under Cooperative Agreement PHY-2019786 (The NSF AI Institute for Artificial Intelligence and Fundamental Interactions, http://iaifi.org/).
JZ acknowledges support by a Project
Grant from the Icelandic Research Fund (grant number 206930).
Simulations and analysis were done on University of Florida's HiPerGator.

We made use of the following software for the analysis:
\begin{itemize}
	\item {\textsc{Python}}: \citet{vanRossum1995}
	\item {\textsc{Matplotlib}}: \citet{Hunter2007}
	\item {\textsc{SciPy}}: \citet{Virtanen2020}
	\item {\textsc{NumPy}}: \citet{Harris2020}
	\item {\textsc{Astropy}}: \citet{Astropy2013,Astropy2018}
	\item {\textsc{SwiftSimIO}}: \citet{Borrow2020, Borrow2021}
\end{itemize}

\section*{Data availability}
Data is available upon request.

\DeclareRobustCommand{\VAN}[3]{#3}

\bibliographystyle{mnras}
\bibliography{bibliography}

\appendix

\section{Numerical convergence}
\label{app:resolution}

\begin{table}
    \centering
    \begin{tabular}{c|cccc}
          Resolution
          & $M/\rm{M}_{\odot}$ & $N$ & $\epsilon_{\rm com}$ & $\epsilon_{\rm max}$ \\
         \hline
         low
             & $1.41\times10^7$ & $890,400$ & 1.22 kpc  & 0.61 kpc \\
        mid 
             & $1.76\times10^6$ & $7,041,720$ & 0.61 kpc & 0.305 kpc \\
        high
             & $2.20\times10^5$ & $55,717,200$ & 0.305 kpc & 0.153 kpc \\
    \end{tabular}
    \caption{Properties of each resolution run for zoom particles: Number $N$ of particles with mass $M$ and comoving softening length $\epsilon_{\rm co}$ that is not allowed larger than the maximum physical softening length $\epsilon_{\rm max}$.  The high resolution level is the version used for our main results.}
    \label{tab:sim_resolutions}
\end{table}

To check the numerical convergence of our results, we run three resolution levels of each of our simulation models.
We summarise the parameters of each resolution level in Table \ref{tab:sim_resolutions}.
In addition to the simulations presented in the main text, the highest level resolution version, we run two lower resolution sets of simulations.
In this Section, we show results for the low, medium and high resolution runs for the CDM and Endothermic models as examples.
The results for the additional models are similar, but we leave them off the plots for clarity.

The medium resolution runs have $7,041,720$ zoom dark matter particles of mass $1.76\times10^6\ \rm{M}_{\odot}$ with a comoving plummer softening length of $0.61$ kpc and a maximum physical comoving length of $0.305$ kpc.
There are
$559,385$ background particles of mass $1.41\times10^7\ \rm{M}_{\odot}$, $223,154$ of mass $1.13\times10^8\ \rm{M}_{\odot}$ and $134,167,340$ of mass $9.03\times10^8\ \rm{M}_{\odot}$.
The background particles have comoving and maximum physical plummer softening lengths of $3.0$ kpc.

The low resolution runs have $890,400$ zoom dark matter particle of mass of $1.41\times10^7\ \rm{M}_{\odot}$ with a comoving plummer softening length of $1.22$ kpc and a maximum physical comoving length of $0.61$ kpc.
There are $165,900$ background particles of mass $1.13\times10^8\ \rm{M}_{\odot}$, $90,150$ of mass $9.03\times10^8\ \rm{M}_{\odot}$ and $167,616$ of mass $7.22\times10^9\ \rm{M}_{\odot}$.
The background particles have comoving and maximum physical plummer softening lengths of $3.0$ kpc.

\begin{figure}
    \includegraphics[width=\linewidth]{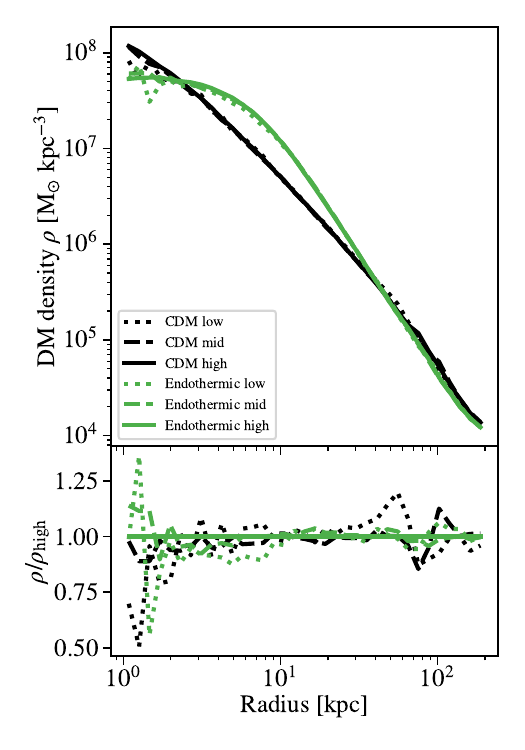}
    \caption{Dark matter density profiles for the main halo for the low (dotted), medium (dashed), and high (solid) resolution runs for the CDM (black) and Born (green) models.  There is not a significant difference in the results between resolution levels.}
    \label{fig:res_density_profiles}
\end{figure}

To emphasise that the main halo is similar across resolution levels, we show in the top panel of Figure \ref{fig:res_density_profiles} the main halo density profiles for the each resolution level (low, medium and high resolutions shown with solid, dashed and dotted lines) for each model (CDM and Born in black and green).
The difference in profile shape remains apparent for all resolution levels, and the lines for each resolution level are consistent for a given model across all radii.
The bottom panel shows the fractional difference of each model compared to the high resolution run of the same model.

\begin{figure}
    \includegraphics[width=\linewidth]{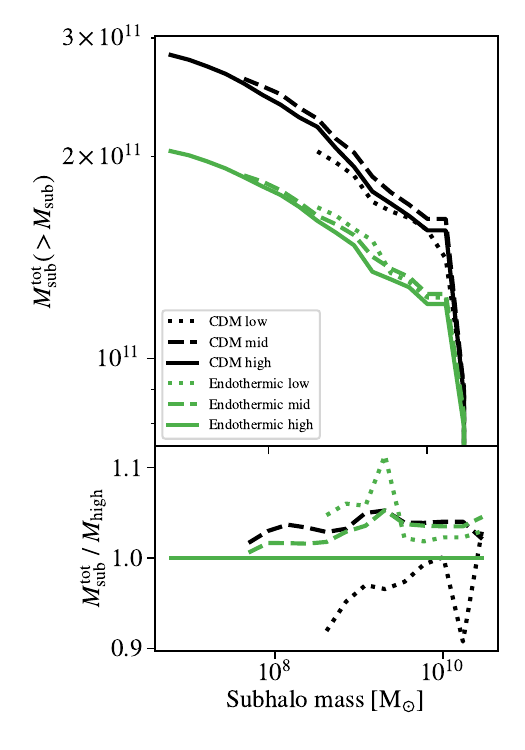}
    \caption{Total mass contained within main halo satellites above a given mass for the low (dotted), medium (dashed) and high (solid) resolution runs for the CDM (black) and Born (green) models.  The higher resolution runs are able to identify less massive substructures.  The differences between the models remain similar across all three resolutions.}
    \label{fig:res_subhalo_mass_functions}
\end{figure}

The difference in subhaloes between the resolution levels is apparent in Figure \ref{fig:res_subhalo_mass_functions}, which shows the total mass contained within subhaloes above a given mass for each run.
Black lines show the CDM model and green lines show the Born model, while the solid, dashed and dotted lines show the low, medium and high resolution runs respectively.
The high resolution lines extend to much lower subhalo masses.
While the low resolution runs only resolve subhaloes down to masses of a few times $10^8\rm{M}_{\odot}$, the medium and high resolution runs resolve subhaloes of approximately $10^8\rm{M}_{\odot}$ and $10^7\rm{M}_{\odot}$ respectively.
This is expected since the lower mass resolution allows lower mass substructure to be identified.

\label{lastpage}

\end{document}